\documentclass[useAMS,usenatbib]{mn2e}
\usepackage{times}
\usepackage{graphics,epsf}
\usepackage{amsmath}                
\usepackage{amsfonts}               
\usepackage{amssymb}                
\usepackage{epsfig}                 
\usepackage{epstopdf}
\usepackage{rotating}
\usepackage{color}
\usepackage{tabularx}
\usepackage{url}
\def \apj{ApJ}
\def \aap{A\&A}
\def \aj{AJ}

\def \mnras{MNRAS}
\def \apjl{ApJ Lett.}

\def \araa{ARA\&A}
\def \na{New A}

\def \memsai{Mem. Societa Astronomica Italiana}
\def \icarus{Icarus}

\begin{document}

\title[Evolution of a stellar disk around a MBH]{Short and long term evolution of a stellar disk around a massive black hole: The role of binaries, the cusp and stellar evolution}

\author[Mikhaloff \& Perets]{Diego N. Mikhaloff \& Hagai B. Perets
\\Physics department, Technion - Israel institute of Technology, Haifa,
Israel 3200002
\\dmikhaloff@tx.technion.ac.il; hperets@physics.technion.ac.il}
\maketitle

\begin{abstract}
We study the dynamical evolution of a stellar disk orbiting a massive
black hole. We explore the role of two-body relaxation, mass segregation,
stellar evolution and binary heating in affecting the disk evolution,
and consider the impact of the nuclear cluster structure and the stellar-disk
mass-function. We use analytic arguments and numerical calculations,
and apply them to study the evolution of a stellar disk (similar to
that observed in the Galactic center; GC), both on the short (few Myr)
and longer (100 Myr) evolutionary timescales. We find the dominant
processes affecting the disk evolution are two-body relaxation and
mass segregation where as binary heating have only a little contribution.
Massive stars play a dominant role in kinematically heating low mass
stars, and driving them to high eccentricities/inclinations. Multi-mass
models with realistic mass-functions for the disk stars show the disk
structure to be mass stratified, with the most massive stars residing
in thinner structures. Stellar evolution plays an important role in
decreasing the number of massive stars with time, thereby leading
to slower relaxation, where the remnant compact objects of these stars
are excited to higher eccentricities/inclinations. At these later
evolutionary stages dynamical heating by the nuclear cluster plays
a progressively more important role. We conclude that the
high eccentricities and high inclination observed for the majority
on the young O-stars in the Galactic Center suggest that the disk
stars had been formed with initially high eccentricities, or that
collective or secular processes (not explored here) dominate the disk
evolution. The latter processes are less likely to produce mass stratification
in the disk; detailed study of the mass-dependent kinematic properties
of the disk stars could therefore provide a handle on the processes
that dominate its evolution. Finally, we find that the disk structure is expected to 
keep its coherency, and be observed as a relatively thin disk even after 100 Myrs; 
two-body relaxation is too inefficient for the disk to assimilate into the 
nuclear cluster on such timescales. It therefore suggests earlier disks 
now containing only older, lower mass stars might still be observed 
in the Galactic center, unless destroyed/smeared by other non-two-body 
relaxation processes. 
\end{abstract}
\begin{keywords}
Galaxy: centre -- galaxies: nuclei -- Galaxy: structure -- stars: kinematics and dynamics -- Galaxy: nucleus
\end{keywords}

\section{INTRODUCTION}

Stellar disks are known to exist around massive black holes (MBHs)
in galactic nuclei, and are typically composed of relatively young
stellar populations. Observations of the central parsec of the Milky-Way
Galaxy show more than 100 young WR, O and B stars \citep{2006ApJ...643.1011P},
likely formed $\sim4-7$ $\mbox{Myr}$ ago. These stars are found
between $\sim0.04-0.5$ pc from the MBH, with $\sim20\%$ of which
reside in a disk-like structure, and are observed to have mean high
eccentricity of $\left\langle e\right\rangle \sim0.3-0.4$. Stellar
disks might therefore be a frequent phenomena in galactic nuclei.
Such disks are thought to form following the infall of gaseous material,
and its formation of a gaseous disk. Such a disk can later fragment
to form a stellar disk \citep[and references therein]{hob+09}. Here
we explore the dynamical evolution of such stellar disks, and focus
on disks similar to that observed in the Galactic center.

\cite{2007ApJ...654..907A} have been the first to describe
the behavior of a stellar disk around a MBH. They used an analytic
approach, and then verified and calibrated them through the use of
N-body simulations. The analytic model they devised includes the effects
of two-body relaxation, and included simple multi-mass models (two
or three populations). Their N-body simulations also included a simplified
stellar evolution. 

Later studies by \citet{2008MNRAS.388L..64C}, \cite{per+08b} and \citet{2009MNRAS.398..429L}
performed additional N-body simulations and considered additional
components and/or more realistic aspects. \citet{2008MNRAS.388L..64C}
included stellar binaries in the disk, the role of a putative intermediate
mass black holes, and initially eccentric disks. \citet{koc+11} discussed
the evolution of the stellar disk, mainly focusing on the effects
of resonant relaxation \citep{1996NewA....1..149R}, not considered
here. The influence of the stellar cusp was studied numerically by \cite{per+08b}
and \citet{2009MNRAS.398..429L}, who performed N-body simulation
for different models (including models with cusp and models with two
stellar disks); they concluded that the cusp could provide an additional
heating of the stellar disk, leading to faster eccentricity/inclination
growth. 

Here we follow-up on the initial analytic work by \cite{2007ApJ...654..907A},
and extend it to include the role of a realistic detailed mass-function,
the effects of stellar evolution, the impact of binary-heating and
additional heating by a stellar cusp. This approach allows us to better
understand and identify the role played by each of these processes and
components, as well as to study the long-term evolution of realistic
disks, which is more difficult (and computationally expensive) to
study using full N-body simulations. 

We begin by describing the dynamical processes that contribute to
the stellar disk evolution, including stellar scattering, mass segregation
and binary-heating (section 2). We then briefly review the structural
components of the nuclear stellar cluster in which the disk is embedded,
and consider various models for the mass function of the stellar disk,
and the possible core-like or cuspy structure of the nuclear stellar cluster (NSC). 
The physical processes we consider as well as the NSC structural components are
used to devise a multitude of possible models for the stellar disk
evolution under diverse conditions, including considerations of stellar
evolution (section \ref{sec:MODELS}). We present the results of the
short and long term evolution of the disks considered in these models
in section \ref{sec:RESULTS:}, and then discuss the results and their
implications (section \ref{sec:DISCUSSION:}). We also briefly mention
other possible physical processes not included in our current modeling.
Finally we conclude and summarize our results in section \ref{sec:SUMMARY}.

\section{The dynamics and physical processes in a stellar disk}
\label{sec:The-dynamics-and}

In the following we briefly review several dynamical processes that
contribute to the disk evolution. We first follow \citet{2007ApJ...654..907A}
in describing the two-body relaxation and mass-segregation (dynamical
friction) of multiple stellar populations (extended to an arbitrary
number of stellar populations), and then consider the analytic description
of the effects of binary heating on stellar disks, not discussed before
in this context.

\subsection{Relaxation and single-star scattering}
\label{sub:Relaxation-and-single-star}

Let us follow \citet{2007ApJ...654..907A} and consider a super massive
black hole (MBH) with mass $M_{\bullet}$ and $N$ single stars with
mass $M$ orbiting the MBH in a initially thin stellar disk, with an
external radius $R_{0}+\Delta R/2$ and an internal radius
$R_{0}-\Delta R/2$. 

When $M_{\bullet}\gg NM$ the relaxation time of such a system is
given by \citet{2008gady.book.....B}.

\begin{equation}
t_{relax}=C\frac{\sigma^{3}}{G^{2}M\rho\ln\left(\Lambda\right)}\label{eq:t_relax}
\end{equation}

With $\rho$ the stellar density and $C\sim1$ (depending only on
the system geometry; e.g. for a spherical case $C\simeq0.34$ \citep{2008gady.book.....B}.
The disk density is given by

\begin{equation}
\rho=\frac{NM}{2\pi R_{0}H\Delta R},\label{eq:rho}
\end{equation}
where $H=\sigma/\Omega$ is the scale height of the disk. We then
obtain 

\begin{equation}
t_{relax}=C\frac{R_{0}2\Delta R\sigma^{4}}{G^{2}NM^{2}\ln\left(\Lambda\right)}.\label{eq:t_relax2}
\end{equation}

Using this Eq. and defining $t_{orb}=2\pi/\Omega$ we can describe
the evolution of the velocity dispersion of disk stars: 

\begin{equation}
\frac{d\sigma}{dt}=\frac{G^{2}NM^{2}\ln\left(\Lambda\right)}{C_{1}t_{orb}R_{0}\Delta R\sigma^{3}},\label{eq:dsigma_dt}
\end{equation}
with $C_{1}=2C$. 

In a disk of stars orbiting an MBH, the root mean square (RMS) eccentricity of the stars
can be related to their velocity dispersion by \citep[and references therein]{2007ApJ...654..907A}
\[
\left\langle e\right\rangle =e_{rms}=\sqrt{2}\frac{\sigma}{v_{K}},
\]
where $v_{K}=(GM_{\bullet}/R_{0})^{1/2}$ is the Keplerian orbital
speed. The RMS inclination of stars is expected to be of the order
of half the RMS eccentricity value. For comparison with observations
of the eccentricities of stars in the Galactic center we therefore
present our results of the RMS eccentricity (and inclination in some
cases) rather than the velocity dispersion. 

\begin{figure}
\includegraphics[scale=0.4]{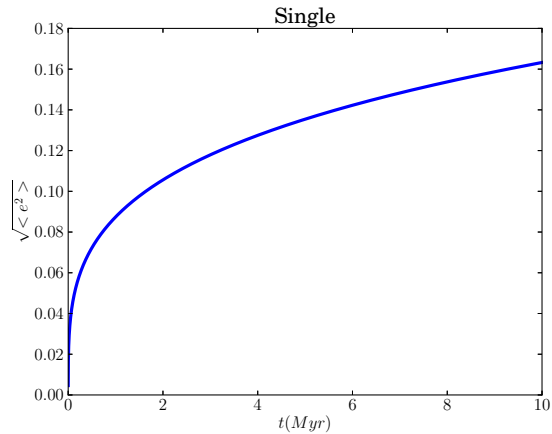}

\caption{\label{fig:sigma-evol-simple}Evolution of the velocity dispersion
of stars in a stellar disk composed of single mass stars orbiting
a MBH; similar to \citet{2007ApJ...654..907A}.}
\end{figure}

Fig. \ref{fig:sigma-evol-simple} shows the evolution of the RMS eccentricity of
stars in the stellar disk, using a similar setup as used by \citet{2007ApJ...654..907A}

\subsection{Multi-mass components and mass segregation}

Stellar scattering had also been considered for multi-mass stellar
populations in the dispersion dominated regime \citep{2007ApJ...654..907A}
. In this regime, for two mass populations, \citet{2004ARA&A..42..549G}
suggest that the treatment of the problem only depends on the amplitude
of the velocity dispersion ($\sigma_{2}$) of the lighter objects,
with mass $M_{2}$, compared with the Hill velocity of the more massive
components , with mass $M_{1}$. The Hill velocity for the more massive
stars is defined as 

\begin{equation}
v_{H,1}=\Omega R_{H,1},\label{eq:v_hill}
\end{equation}
where $R_{H,1}=R_{0}(M_{1}/M_{2}){}^{1/3}$. The dispersion
dominated regime occurs when $\sigma_{2}>v_{H,1}$ . In this regime
the speed of the single stars is similar to the velocity dispersion
and scattering encounters are well approximated by two body dynamics.
In the shear dominated regime, not discussed here, it is necessary
to take into account the tidal gravity of the central black hole.

When the velocity dispersion of the low-mass stars satisfies $v_{esc,1}>\sigma_{2}>v_{H,1}$,
with $v_{esc,1}$ the escape velocity from massive stars, the exchange
of momentum between the heavier and lighter bodies occurs through collisionless
gravitational deflections, where gravitational focusing is important
\citep{2004ARA&A..42..549G}.

Considering only two stellar populations composed of $N_{1}$ and
$N_{2}$ stars with stellar masses $M_{1}$ and $M_{2}$, and velocity
dispersions $\sigma_{1}$ and $\sigma_{2}$, respectively, we obtain
\citep{2007ApJ...654..907A}: 

\begin{equation}
\frac{d\sigma_{1}}{dt}=\frac{N_{1}M_{1}^{2}}{A_{1}t_{orb}\sigma_{1}^{3}}\ln\left(\Lambda_{1}\right)-\frac{N_{2}M_{1}M_{2}\ln\left(\Lambda_{12}\right)}{A_{2}t_{orb}}\frac{\sigma_{1}}{\overline{\sigma}_{12}^{4}}\left(1-\frac{E_{2}}{E_{1}}\right)\label{eq:dsigma1_dt}
\end{equation}

\begin{equation}
\frac{d\sigma_{2}}{dt}=\frac{N_{2}M_{2}^{2}}{A_{1}t_{orb}\sigma_{1}^{3}}\ln\left(\Lambda_{2}\right)-\frac{N_{1}M_{1}M_{2}\ln\left(\Lambda_{12}\right)}{A_{2}t_{orb}}\frac{\sigma_{2}}{\overline{\sigma}_{12}^{4}}\left(1-\frac{E_{1}}{E_{2}}\right),\label{eq:dsigma2_dt}
\end{equation}
where $\overline{\sigma}_{12}=(\sigma_{1}+\sigma_{2})/2$ and
$A_{i}=(C_{i}R_{0}\Delta R)/G^{2}$ for $i=1$ and 2, and $\Lambda_{1}\,\Lambda_{2}$
are the appropriate Coulomb logarithms for the two stellar populations,which
depend on the respective scale height of the disk. $\Lambda_{12}$
is related to the scale height at which interactions between
the two stellar species occur \citep[see][for further details]{2007ApJ...654..907A}
The constant $C_{2}$ is related to the constant $C_{1}$, and also
depends on the geometry of the system. i.e for a three dimensional
system $C_{1}/C_{2}\simeq3.5$ (for a detailed analysis please
see at \citealp{1988Icar...74..542S}); here we adopt the same value.

We can extend such a model to an arbitrary number of different stellar
species, $N$. Each species adds an additional coupled equation and
contributes a coupling term to each of the other equations. For the
general case the time evolution of the velocity dispersions for each
of the stellar species is given by a set of coupled equations

\begin{equation}
\frac{d\sigma_{j}}{dt}=\frac{N_{j}M_{j}^{2}}{A_{1}t_{orb}\sigma_{j}^{3}}\ln\left(\Lambda_{j}\right)-\sum_{k\neq j}\frac{N_{k}M_{j}M_{k}\ln\left(\Lambda_{jk}\right)}{A_{2}t_{orb}}\frac{\sigma_{j}}{\overline{\sigma}_{jk}^{4}}\left(1-\frac{E_{k}}{E_{j}}\right),\label{eq:dsigmaj_dt}
\end{equation}
for $i=1,2,\dots,N$ and $k=1,2,\dots,N$ (with $k\neq j$). 

From analyzing these set of equations we note that for two species,
$k$ and $j$ for which $E_{k}>E_{j}$, the population of $k-$species particles dynamically
heat the population of the $j$-species ones (leading to increased velocity
dispersion) and vice versa (a less energetic population dynamically
cools a more energetic one, dumping its velocity dispersion).

\subsection{Interaction between the stellar disk and the nuclear cluster}

The two-body relaxation of disk stars due to interaction with the
background stars of the nuclear cluster can be modeled in a similar
method to the two-body relaxation by different stellar populations
discussed above. In this case, the cluster stellar population is considered
as a distinct population with different properties. The disk stars
interact with those cluster stars which go through the disk, i.e if the
volume of the disk is $V=4\pi R_{0}H\Delta R$ , the number of clusters
stars considered is the nuclear cluster stellar density times this
volume. The velocity dispersion of this isotropic cluster population
is $\sigma_{cluster}=V_{K}/\sqrt{3}$ \citep{2008gady.book.....B}
with $V_{K}$ the keplerian velocity around the MBH. We now use the
same type of two-population interaction term in Eq. \ref{eq:dsigmaj_dt}
to account for the effects of the cluster stars on the dis stellar
population. Note that the cluster stellar population is much greater
than the disk population, and can be assumed as a constant ``heat''
bath.

\subsection{Binary Heating}
\label{sub:Binary-Heating}

Any stellar population is likely to contain some fraction of binary
stars. An encounter between single stars and binaries could lead to
an energy exchange between the binary orbital energy and the kinetic
energy of the stars, i.e., energy can be exchanged between the inner
degree of freedom of the binary and the outer external kinetic energy
of the stars; thereby binaries can assist in dynamically heating the
stellar disk.

Let us consider a binary system composed of stars with masses $m_{1}$
and $m_{2}$, center of mass velocity $V_{cm}$ , and orbital velocity
$V$ with respect to the center of mass of the system (we assume a
circular binary for simplicity). The total energy of the binary can
be expressed as \citep{2008gady.book.....B}:

\[
E=\frac{1}{2}\mu V-G\frac{m_{1}m_{2}}{r}+\frac{m_{1}+m_{2}}{2}V_{cm}^{2},
\]
where $\mu=\frac{m_{1}m_{2}}{m_{1}+m_{2}}$ is the reduced mass. The
sum of the first two left element define the binding energy, $E_{b}$,
in terms of the semi-major axis 

\begin{equation}
E_{b}=-G\frac{m_{1}m_{2}}{2a}.\label{eq:E_b}
\end{equation}

Binary-single encounters had been explored extensively, with the pioneering
studies by \citet{1975MNRAS.173..729H} and \citet{1975AJ.....80..809H}.
Binaries can be divided into two dynamical categories; ``Hard Binaries''
and ``Soft Binaries'': when the kinetic energy of the intruding
mass, $E_{k}$, is such that $E_{k}/m_{b}V^{2}\ll1$, (where $m_{b}=m_{1}+m_{2}$
is the binary total mass)a binary is considered to be hard (where
the lower the ratio the harder the binary); it is considered to be
a soft binary when $E_{k}/m_{b}V^{2}\gg1$.

\citet{1975MNRAS.173..729H} and \citet{1975AJ.....80..809H} found
that following a binary-single encounter soft binaries become softer
and hard binaries become harder (on average) , a behavior typical
termed as ``Heggie's law''. 

The long term typical evolution of a binary randomly encountering
stars in some dense environment can then be modeled using a simple
approach. Considering a binary in a field of incoming stars with typical
mass $m$ and particle density $n$ , the averaged evolution of the
binding energy of the binary due to many encounters is given by

\begin{equation}
\left\langle \frac{dE_{b}}{dt}\right\rangle =n\left\langle \sigma_{E}V\right\rangle E_{b},\label{eq:dE_b_dt}
\end{equation}
where $V$ the typical relative velocity of the intruders, and $\sigma_{E}$
the cross section of the single-binary collisions. We may now consider
the contribution of a stellar population $i$ with stellar density
$n_{i}$ which is composed of stars with typical mass $m_{i}$, to
the binary evolution; when the velocity distribution of incoming stars
is Maxwellian and taking Eq. \ref{eq:dE_b_dt} can be expressed
as \citep{1992AJ....103.1955H}:

\begin{equation}
\left\langle \frac{dE_{b}}{dt}\right\rangle =\frac{\sigma_{E}^{*}\left(6\pi\right)^{\frac{1}{2}}G^{2}m_{b}^{2}\rho_{i}}{4\sqrt{\left\langle V_{i}^{2}\right\rangle }}\left(\frac{m_{i}+m_{b}}{m_{b}}\right)^{1.2}\left(\frac{1}{2\frac{m_{i}}{m_{b}}+\frac{1}{3}}\right),\label{eq:dE_v_dt_f}
\end{equation}
where $m_{b}=m_{1}+m_{2}$ , $\rho_{i}=n_{i}m_{i}$. \citet{1992AJ....103.1955H}
made use of N-Body simulations to study the dependence of $\sigma_{E}^{*}$
on the intruder mass ($m_{i}$) and the binary mass ($m_{b}$) in
the mass-ratio range $2m_{i}/m_{b}\in\left[0.01,10000\right]$ and
found that the the empirical functional dependence can be formulated
by 

\[
\sigma_{E}^{*}=\frac{a+cx+ex^{2}+gx^{3}}{1+bx+dx^{2}+fx^{3}}
\]

with $x=\ln\left(2m_{i}/m_{b}\right)$, $a=1.26$, $b=-0.15$, $c=0.054$,
$d=0.02$, $e=0.1$, $f=0.0045$ and $g=0.015$.

For a single star the average kinetic energy is $\left\langle E\right\rangle =3m\sigma_{i}^{2}/2$
with $\sigma_{i}$ the velocity dispersion. Taking $\sigma_{i}\sim\sqrt{\left\langle V_{i}^{2}\right\rangle }$
(encounter velocity $\sim$dispersion velocity) and applying energy
conservation for a system of binary stars with density $n_{b}$we
get

\begin{equation}
n_{i}\left\langle \frac{dE}{dt}\right\rangle =-n_{b}\left\langle \frac{dE_{b}}{dt}\right\rangle .\label{eq:equal_energy}
\end{equation}

Eqs. \ref{eq:dE_v_dt_f} and \ref{eq:equal_energy} give

\begin{equation}
\frac{d\sigma_{i}}{dt}=D_{i}\frac{n_{b}}{3n_{i}m_{i}\sigma_{i}^{2}}\label{eq:binary_heting}
\end{equation}

with 

\begin{equation}
D_{i}=\frac{\sigma_{E}^{*}\left(6\pi\right)^{\frac{1}{2}}G^{2}m_{b}^{2}\rho_{i}}{4}\left(\frac{m_{i}+m_{b}}{m_{b}}\right)^{1.2}\left(\frac{1}{2\frac{m_{i}}{m_{b}}+\frac{1}{3}}\right).\label{eq:D_constant}
\end{equation}

We can now use Eq. \ref{eq:binary_heting} as the binary-heating term
to be added to the dynamical evolution equation for the stellar disk
populations (Eq. \ref{eq:dsigma_dt}); allowing us to model the binary-heating
contribution. 

We are now in a position to asses whether binary heating can play
an important role in heating the stellar disk.. For the simplest
case of a disk with only one population of same-mass stars and a binary
fraction $n_{b}$, the velocity dispersion evolution equation is obtained
by summing the scattering term and the binary heating to get (taking
$m_{b}=m_{i}=M$)

\[
\frac{d\sigma}{dt}=\frac{G^{2}NM^{2}\ln\left(\Lambda\right)}{C_{1}t_{orb}R_{0}\Delta R\sigma^{3}}+D_{i}\frac{n_{b}}{3n_{i}M\sigma^{2}},
\]
 or after some math

\begin{equation}
\frac{d\sigma}{dt}=\frac{G^{2}NM^{2}}{t_{orb}R_{0}\Delta R\sigma^{3}}\left[\frac{\ln\left(\Lambda\right)}{C_{1}}+\frac{N_{b}}{N}K\right],\label{eq:bin-vs-single}
\end{equation}
where $N_{b}$ is the number of binary stars, and 

\[
K=\frac{\sigma_{E}^{*}\left(6\pi\right)^{\frac{1}{2}}2^{-0.8}}{7}\simeq0.53.
\]

\begin{figure}
\begin{centering}
\includegraphics[scale=0.4]{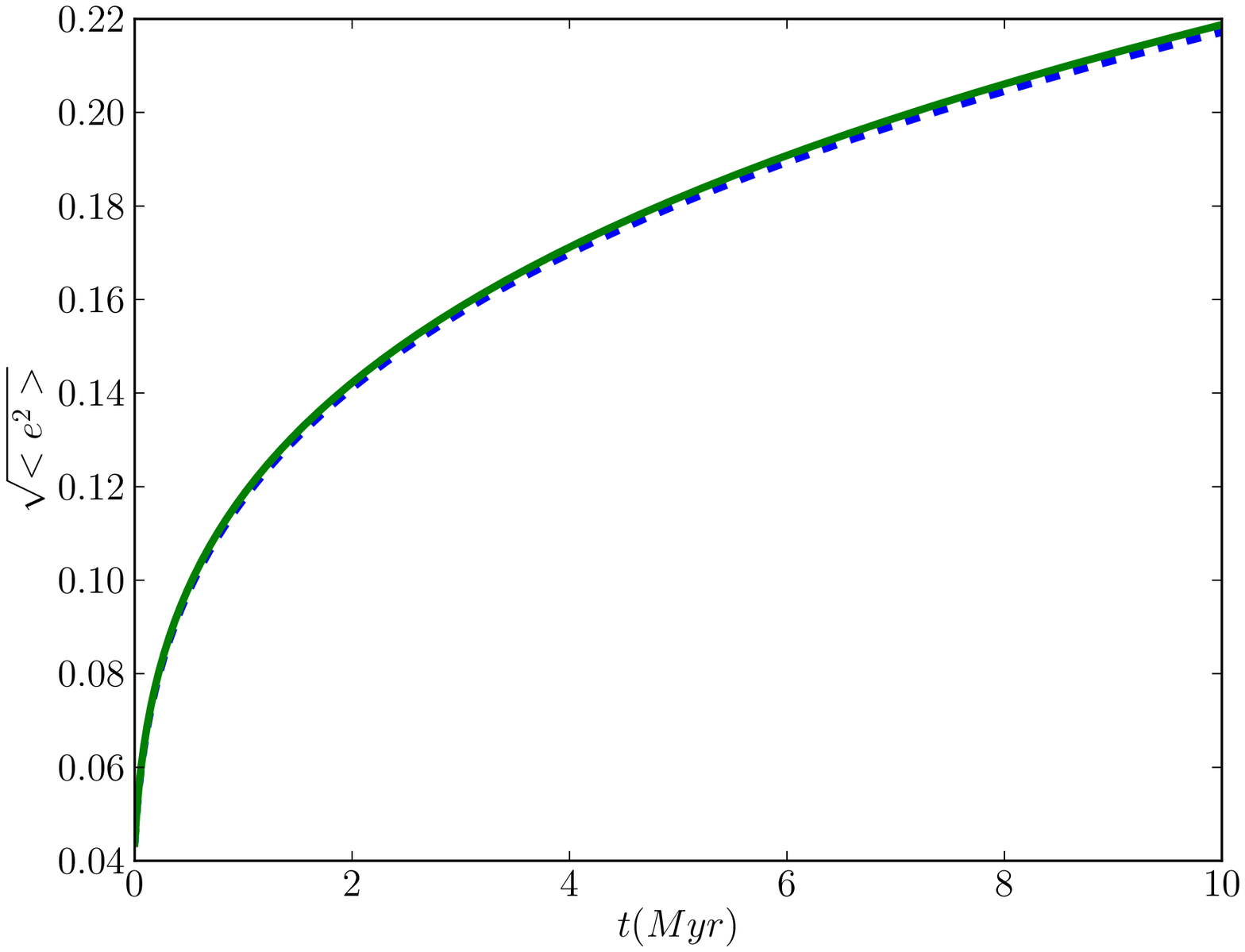}
\includegraphics[scale=0.4]{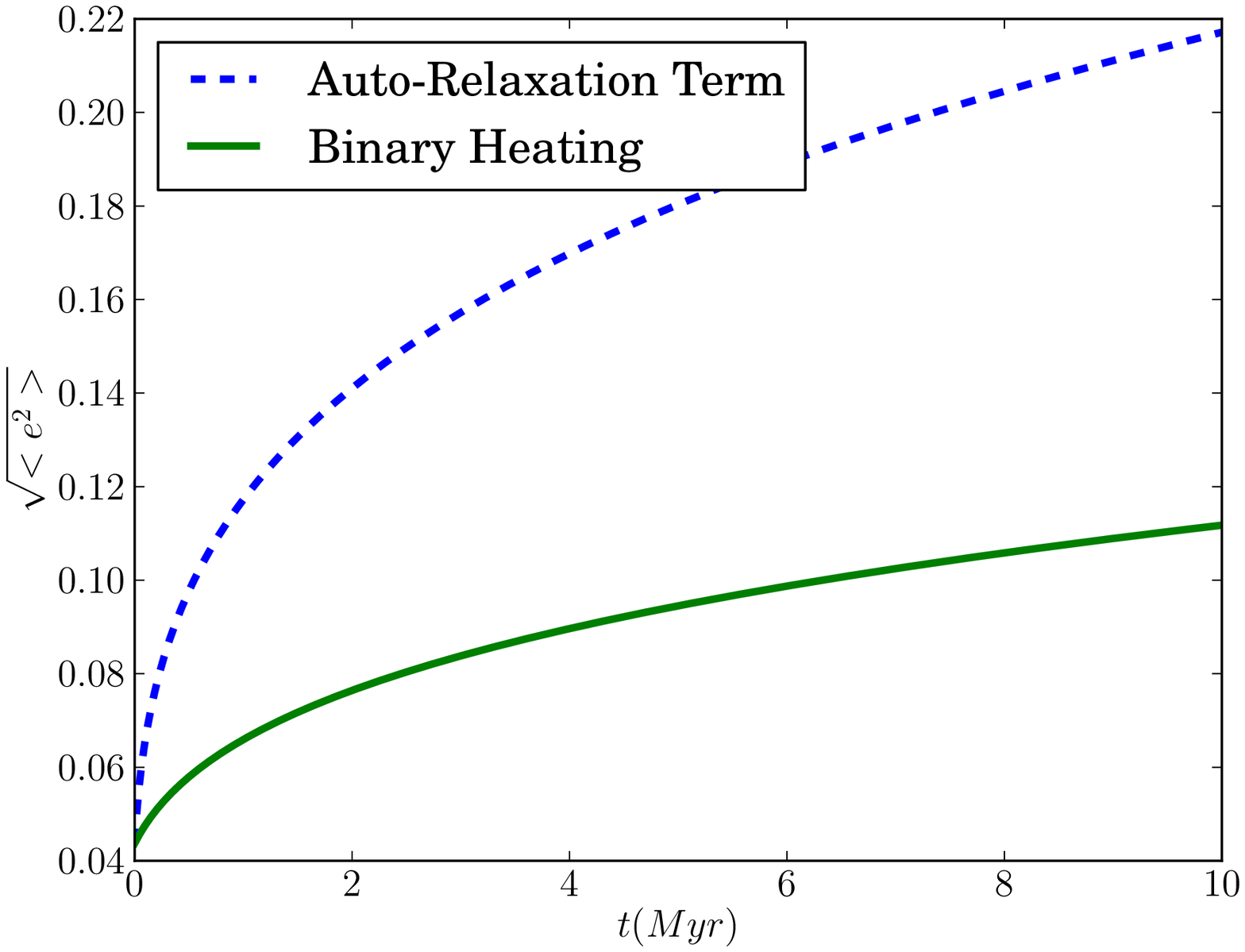}
\caption{Evolution of RMS eccentricity for model X (in Section \ref{sec:MODELS}
and Table \ref{tab:models} for more information about the models).
\label{fig:binaries-2}. In the top panel is plotted the evolution
of RMS eccentricity through time, the dashed line is the model only
considering auto-relaxation and the solid one is when binary heating
is included. In the bottom panel are ploted the the evolution of the
RMS eccentricity when only auto-relaxation (dashed line) or binary
heating is considered (solid line).}
\end{centering}
\end{figure}

We can now compare the relative contribution by single-star scattering
and binary heating through the comparison of the two terms in Eq. \ref{eq:bin-vs-single}.
If we consider some typical initial conditions in the Galactic center, taking 
$\ln\left(\Lambda\right)\simeq7.5$ (as the disk heats-up the scale height grows but with only a small
corresponding change in $\ln(\Lambda)$ due to the logarithmic dependence)
, $C_{1}\simeq3$ and a binary fraction of 2/3, $N_{b}=(2/3)N$,  we 
find that the relative contribution of binary heating is only one tenth
of the contribution due to single star scattering. This can be
seen explicitly in the evolution of the velocity dispersion shown
in Fig. \ref{fig:binaries-2}. The left panel shows how the exclusion of the
binary heating contribution (for a disk in which $M_{single}=M_{binary}=20M_{\odot}$)
makes only a very small difference (see also the effect of
accounting only for binary heating). We can therefore conclude that
although the binaries binding energy could be larger that the kinetic
energy of the disk, the rate of extraction of this energy and its
exchange into the disk kinetic energy is negligible, at least for
the relatively thin disks considered here.

\subsection{Full model}

We can combine the stellar disk contributions described above to obtain
the full set of equations describing our model, including multi-mass
populations, binaries and nuclear cluster stellar population. This
is summarized below;

\begin{eqnarray}
\frac{d\sigma_{j}}{dt} & = & \frac{N_{j}M_{j}^{2}}{A_{1}t_{orb}\sigma_{j}^{3}}\ln\left(\Lambda_{j}\right)\nonumber \\
 &  & -\sum_{k\neq j}\frac{N_{k}M_{j}M_{k}\ln\left(\Lambda_{jk}\right)}{A_{2}t_{orb}}\frac{\sigma_{j}}{\overline{\sigma}_{jk}^{4}}\left(1-\frac{E_{k}}{E_{j}}\right)\nonumber\\
 &  & +\sum_{i}D_{ij}\frac{n_{binary-i}}{3n_{j}M_{j}\sigma_{j}^{2}}\\
 &  & -\sum_{i}\frac{N_{binary-i}M_{j}m_{binary-i}\ln\left(\Lambda_{j-binary-i}\right)}{A_{2}t_{orb}}\cdot\nonumber \\
 &  & \frac{\sigma_{j}}{\overline{\sigma}_{j-binary-i}^{4}}\left(1-\frac{E_{binary-i}}{E_{j}}\right)\nonumber\\
 &  & -\frac{N_{cusp}M_{j}M_{cusp}\ln\left(\Lambda_{j-cusp}\right)}{A_{2}t_{orb}}\frac{\sigma_{j}}{\overline{\sigma}_{j-cusp}^{4}}\left(1-\frac{E_{cusp}}{E_{j}}\right)\nonumber
\end{eqnarray}
where 
\begin{enumerate}
\item $\Lambda_{j-cusp}=\overline{\sigma}_{j-cusp}\Delta R/2GM_{\odot}$
\item $\overline{\sigma}_{j-cusp}=(\sigma_{cusp}+\sigma_{j})/2$, 
\item $\sigma_{cusp}=\frac{1}{3}\sqrt{GM_{T}\left(R\right)/R}$ 
\item $M_{T}\left(R\right)=M_{cusp}\left(R\right)+M_{SMBH}$ 
\item $M_{cusp}\left(R\right)=\int_{R_{min}}^{R_{max}}4\pi r^{2}\rho_{cusp}\left(r\right)dr$
, where we consider the effective region to affect the stars to be
between $R_{min}=0.5R$ and$R_{max}=1.5R$ and 
\item $\rho_{cusp}\left(r\right)\propto r^{-\gamma}$ 
\item $\Lambda_{j-binary-i}=\frac{\left(\frac{\sigma_{j}+V_{i-orbital}}{2}\right)^{2}}{2GM_{binary-i}}$,
where $V_{i-orbital}^{2}=\frac{Gm_{binary-i}}{a_{i}}$ and $a_{i}$
is semi-major axis of the binary.
\item $\Lambda_{jk}=\frac{\overline{\sigma}_{jk}\Delta R}{2GM_{k}}$ 
\end{enumerate}

\section{Structural components of the nuclear stellar cluster}

In the following we describe the observed properties of the cluster
and the embedded stellar disk, and the models we use in our analysis.

\subsection{Observed disk structure and mass function}
\label{sub:Disk-strcuturestructure-and}

Beginning with the analysis of \citet{2003ApJ...590L..33L}, observations
have shown the existence of a young stellar disk near the MBH in the
Galactic center. \citet{2006ApJ...643.1011P} showed that many of the
observed young O/WR/B-stars reside in a disk-like structure between
$\sim0.05-0.5$ pc from the MBH. 

Early measurements suggested the disk-stars revolve around the MBH
in somewhat eccentric orbits.\cite{2009ApJ...697.1741B} found an RMS
eccentricity of $\left\langle e\right\rangle =0.37\pm0.07$ for the disk stars,
but the most recent studies suggest a smaller value of $\left\langle e\right\rangle =0.27\pm0.07$ 
\citep{2014ApJ...783..131Y}.
The disk-stars appear to be young, forming only $4-7$ Myrs ago \citep{2006ApJ...643.1011P}. The
radial surface density profile of the disk can be described by a power-law
distribution, $\Sigma\left(R\right)\sim R^{-\beta}$, with $\beta$
in the range $\left[1.5,2.3\right]$ \citep{2006ApJ...643.1011P,2009ApJ...690.1463L,2009ApJ...697.1741B,2010ApJ...708..834B}. 

Initially it was thought that $\sim40-50\%$ of the young stars belong
in the disk structure; however, more recent studies suggest that only
$\sim20\%$ of these reside in the disk, while most of the stars have
more spherical distribution, outside the disk \citep{2009ApJ...690.1463L}. 
Here we consider only the evolution of disk stars.

In most galactic environments the initial mass function (IMF) describes
the distribution of stellar masses at birth. Analysis of observations
(\citealt{1955ApJ...121..161S}, \citealt{2001ApJ...554.1274C}, \citealt{2002Sci...295...82K})
shows an almost universal power law behavior, with $dN\left(m\right)\propto m^{-\alpha\left(m\right)}dm$,
with a typical exponent of $\alpha\sim2.3$, for stars above one Solar
mass and up to $\sim120$ $M_{\odot}$. The analysis done by \citet{2010ApJ...708..834B}
based on spectroscopic survey made with SINFONI conclude that the
present day mass function and the IMF for the disk of young stars
in the Galactic center obey a power law, but the slope is flatter
than the ``regular'' IMF. In particular, they conclude that $\alpha=0.45\pm0.3$.
In such a ``top heavy'' IMF, massive stars are much more frequent
compared with the Salpeter IMF.

\subsection{Disk models}

In our models we consider an initially thin cool disk ($H\ll\Delta R$
and low initial eccentricities for the star, i.e. we assume an initially
low velocity dispersion for the disk stars) rotating at angular speed
$\Omega=V_{K}/R_{0}$ (with $V_{K}$ the Keplerian velocity).

We only consider the evolution of the disk stars and do not discuss
the origin and evolution of the more spherical component of the young
stars in the Galactic center.

We studied three different types of disks: a single-mass disk, an
Initial Mass Function (IMF) disk and disk with binaries.

\subsubsection{Mass function}

We consider various mass functions for the stars in the stellar disk
and the cusp. These different models are summarized in the following.

\paragraph*{Single-mass model: }

In the simplest, unrealistic model studied here, we consider a stellar
disk composed of single-mass stars, similar to the basic models considered
by \citet{2007ApJ...654..907A}, as a test-case for comparison.

\paragraph*{Top-heavy initial mass-function model:}

In this model we assume the initial mass function of the disk stars
is a top-heavy mass function with masses in the range $0.6-120$ ${\rm M_{\odot}}$,
as suggested by \citet[see discussion above]{2010ApJ...708..834B}.

\paragraph*{Salpeter mass function:}

In this case we assume the initial mass function of the stars in the
stellar disk is a Salpeter mass function with masses in the range
$0.6-120$ ${\rm M_{\odot}}$ .

\subsubsection{Binarity}

Our knowledge about binaries in the GC is currently limited, but Observation by \citealt{1999ApJ...523..248O}
and \citealt{2006ApJ...649L.103M} show the existence of at least some
young binaries in the Galactic center. In particular, they found that
one of the brightest O/WR star in the GC is a massive binary star
($M_{b}\sim100M_{\odot}$). More recent works \citep{2007ApJ...659.1241R}
found a few additional binary O-star candidates, though no binaries
had been found among the young B-stars. These studies either rely
on detection of eclipsing binaries or through radial velocity detection
of short-period binaries, and are therefore mostly sensitive to close
binaries. It is therefore difficult to estimate the underlying binary
fraction among the young stars in the Galactic center using these
limited statistics. Taking them at face value would suggest a very
high binary fraction, comparable or even larger than the high binary
fraction observed in the filed for similarly young massive stars \citep{san+12}.

On the theoretical side, most binaries in the Galactic center are
expected to be ``soft'' binaries and therefore have relatively short
lifetime due to perturbations by other stars \citep{1977ApJ...216..883B,1982A&A...111....1O,hop09,per09,ale+14}, or merge due to secular perturbations by the massive black hole \citep{per09b,ant+12,pro+15}.
The binary fraction among the old stellar population in the nuclear
cluster is likely to be negligible, and we do not include it in any
of our calculations.

Binaries in the disk could, however, be ``hard'' in respect to the
other disk stars, given the low-velocity dispersion in the disk. The
young age of the disk also suggests that most of the initially formed
binaries should still survive. Overall it is therefore likely that
the binary fraction of the massive stars in the Galactic center stellar
disk is high as in the field ($>50\%$), and that the binary fraction
does not evolve much since its formation. 

As we discussed above, binaries are generally not very important for
the disk evolution, but we do consider them in some of our models
for completeness, in which case we assume a binary fraction of $50\%$.

\subsection{Cusp and core models for the nuclear cluster}

A stellar cluster around a massive black hole is expected to evolve
into a stellar cusp structure over a relaxation time \citep{1976ApJ...209..214B,1977ApJ...216..883B}.
For a spherically symmetric distribution of equal-mass stars analytic
considerations and and N-body simulations show that in equilibrium
the density profile of the cluster has a power-law distribution, $\rho\propto r^{-\gamma}$,
with $\gamma=7/4$ \citep{1976ApJ...209..214B}. Mass segregation
in multi-mass clusters produce mass-dependent density profiles for
the different mass stellar populations, with $\rho\left(m\right)\propto r^{-\gamma\left(m\right)}$,
where $\gamma\left(m\right)=1.5+m/(4m_{\max})$, with $m_{\max}$
the mass of the heaviest stellar element \citep{1977ApJ...216..883B}. 

\citet{2006ApJ...645L.133H} used Fokker-planck calculations to show
that $\gamma$ should be $1.4$ for solar mass MS stars and up to
$2$ for stellar black holes (SBH). Similar results from Monte-Carlo
calculations were found by \citet{2006ApJ...649...91F} However, observations
of red-giants in the Galactic center suggest a core-like structure
in the inner regions of the GC ($\gamma$ in the range $0-0.5$; \citealp{2009ApJ...703.1323D,2010ApJ...708..834B,2010RvMP...82.3121G}.
Nevertheless, it is not yet clear whether the distribution of red-giants
reflect the overall distribution of stars in the GC, and various models
had been suggested to explain both a ``real'' core distribution
\citep{mer+10} or an apparent one (i.e., only reflecting the distribution
of red-giants; \citealp{dal+09,ama+14,aha+15}).

In light of the above discussion, we consider two possible models:
(1) A cusp model dominated by a steep power-law distribution of SBHs
($\gamma=2$) and (2) A core model with a shallow power-law distribution
($\gamma=0.5$).

\subsection{Stellar evolution and the long term evolution of disks }

\label{sub:stellar-evolution}

In addition to our study of the role of the different NSC and disk
components, we also considered the effects of stellar evolution, which
become especially important once the long term evolution of the stellar
disk is explored. We considered the evolution of disks with various
properties (see below) both at times comparable to the observed
stellar disk in the GC (up to 10 Myrs), as well as a longer
term (100 Myrs) evolution. 

We first considered simple models in which we did not introduce any
stellar evolution, and assumed stars do not change over time. We then
consider the effects of stellar evolution in a simplified manner,
For each stellar population in a given mass-bin we consider the appropriate
MS lifetime for a star of such mass, $T_{MS}$, according to stellar
evolutionary models. After that period we replace the star in the
model with the stellar remnant it produces, assuming a simplified
prescription as described in table \ref{tab:mass_loss}; in particular
the continuous mass-loss process is simplified and the mass loss is
assumed to be immediate. We only considered evolution of up to 100
Myrs and hence we do not consider the stellar evolution of stars with
masses smaller than 6 M$_{\odot}$. 

\begin{table}
{\tiny }%
\caption{\label{tab:mass_loss} The final mass and final type after $T_{MS}$
(of each star) for different intervals of mass. However star with
initial mass less than $6.5M_{\odot}$ has $T_{MS}$ greater than
100Myr. }

\begin{tabular}{|c|c|c|}
\hline 
Initial Mass $\left[M_{\odot}\right]$ & Final remnant mass (after $T_{MS}$)$\left[M_{\odot}\right]$ & Type of remnant\tabularnewline
\hline 
\hline 
$30<M\leq120$ & $10$ & SBH\tabularnewline
\hline 
$15<M\leq30$ & $7$ & SBH\tabularnewline
\hline 
$8\leq M\leq15$ & $1.4$ & NS\tabularnewline
\hline 
$6\leq M\leq8$ & $1$ & WD\tabularnewline
\hline 
\end{tabular}{\tiny \par}

\end{table}

\section{Detailed models}
\label{sec:MODELS}

In the following we describe the specific evolutionary models we discuss
in detail. We consider a disk of stars orbiting a MBH with mass $M_{\bullet}=4\times10^{6}M_{\odot}$.
The inner disk radius is $R_{in}=0.05$pc, and the external radius
$R_{out}=0.15$ pc (so $\Delta R=0.1$pc). The initial velocity dispersion
is assumed to be low (i.e. a cool thin disk with low eccentricity
orbits) (initial velocity dispersion $\sigma_{j0}\ll V_{K}$ for any
disk population j; we take than $\sigma_{j0}=0.03V_{K}$) . We take
a disk surface density of $\Sigma\left(R\right)\propto R^{-1}$ (following
similar assumptions as \citealp{2008MNRAS.388L..64C} and \citealp{pro+15}). 

We consider several models for the nuclear cluster, both cusp and
core models. We study two different cusp models. The CUSP I model
is composed of single, solar mass stars \citep[see][for details]{pro+15}
with a density profile profile

\begin{equation}
\rho\left(r\right)=\rho_{0}\left(\frac{r}{r_{0}}\right)^{-\gamma}\left[1+\left(\frac{r}{r_{0}}\right)^{2}\right]^{\frac{\left(\gamma-1.8\right)}{2}},\label{eq:cusp_density}
\end{equation}
with $r_{0}=0.5$ pc, $\rho_{0}=5.2\times10^{5}\,{\rm M_{\odot}\, pc^{-3}}$
and $\gamma=2$.

The CUSP II model is a multi-component cusp, with two populations;
MS stars of 1 ${\rm M_{\odot}}$, and SBHs of 10 ${\rm M_{\odot}}$.
These have a power-law $\rho\propto r^{-\gamma}$, distributions with $\gamma_{MS}=1.4$ $\gamma_{SBH}=2$,
respectively; following the results of \citet{2006ApJ...645L.133H}.
Finally the CORE model is similar to CUSP I model but with $\gamma=0.5$.
The models properties, are summarized in Table \ref{cusp}. 

\begin{table}
\caption{\label{cusp} NSC Models.}
\begin{tabular}{|c|c|c|c|c|}
\hline 
Model & Star & $M\left(M_{\odot}\right)$ & $\gamma$ & $\rho_{0}\left(M_{\odot}\mbox{pc}^{-3}\right)$\tabularnewline
\hline 
\hline 
CUSP I & MS & $1$ & $2.0$ & $5.2\times10^{5}$\tabularnewline
\hline 
CUSP II & MS & $1$ & $1.4$ & $1.9\times10^{6}$ \tabularnewline
\hline 
 & SBH & $10$ & $2.0$ & $1.8\times10^{5}$\tabularnewline
\hline 
CORE & MS & $1$ & $0.5$ & $5.2\times10^{5}$\tabularnewline
\hline 
\end{tabular}
\end{table}

We studied different models for various disk and cluster combinations.
A brief summary of the properties of the models is given in Table
\ref{tab:models}, including the disk mass function and its binary
fraction.

\begin{table*}
\caption{\label{tab:models}Summary of the properties of the disk and nuclear cluster models,
including their mass functions and density profiles.}
\begin{tabular}{|c|c|c|c|c|c|c|c|c|}
\hline 
$\#$ & Disk & Cusp/Core & Masses in the cusp & $\gamma$ & $M_{\min}$ & $M_{\max}$ & $\Gamma$ & Binary fraction\tabularnewline
\hline 
\hline 
I & O-stars with $M_{\star}=20M_{\odot}$ & - & - & - & - & - & - & -\tabularnewline
\hline 
II & O-stars with $M_{\star}=20M_{\odot}$ & CUSP I  & $1M_{\odot}$ MS stars  & $1.5$ & - & - & - & -\tabularnewline
\hline 
III & O-stars with $M_{\star}=20M_{\odot}$ & CORE  & $1M_{\odot}$ MS stars  & $0.5$ & - & - & - & -\tabularnewline
\hline 
IV & Salpeter IMF & - & - & - & $0.6M_{\odot}$ & $120M_{\odot}$ & $2.35$ & -\tabularnewline
\hline 
V & Salpeter IMF & CUSP I & $1M_{\odot}$ MS stars  & $1.5$ & $0.6M_{\odot}$ & $120M_{\odot}$ & $2.35$ & -\tabularnewline
\hline 
VI & Salpeter IMF & CORE & $1M_{\odot}$ MS stars  & $0.5$ & $0.6M_{\odot}$ & $120M_{\odot}$ & $2.35$ & -\tabularnewline
\hline 
VII & Top-Heavy IMF & - & - & - & $0.6M_{\odot}$ & $120M_{\odot}$ & $0.45$ & -\tabularnewline
\hline 
VIII & Top-Heavy IMF & CUSP I & $1M_{\odot}$ MS stars & $1.5$ & $0.6M_{\odot}$ & $120M_{\odot}$ & $0.45$ & -\tabularnewline
\hline 
IX & Top-Heavy IMF & CORE & $1M_{\odot}$ MS stars & $0.5$ & $0.6M_{\odot}$ & $120M_{\odot}$ & $0.45$ & -\tabularnewline
\hline 
X & O-stars with $M_{\star}=20M_{\odot}$ & - & - & - & - & - & - & $0.4$\tabularnewline
\hline 
XI & O-stars with $M_{\star}=20M_{\odot}$ & CUSP II  & $1M_{\odot}$ MS /$10M_{\odot}$ SBH & $1.4$/ $2.0$  & - & - & - & -\tabularnewline
\hline 
\end{tabular}

\end{table*}

\section{RESULTS}
\label{sec:RESULTS:}

In the following we present the evolution of stellar disks for the
different models considered. We first study simpler cases with single-mass
disks and then progressively consider more and more realistic cases.
These include disks composed of stars with a range of masses and different
mass-functions, and the effects of different types of NSCs. In this
case we also explore the differential evolution of stars of different
masses in the same disk and their stratification. We then consider
the effects of stellar evolution on the disk, and the long term (100
Myrs) evolution of stellar disks.

\subsection{Evolution of Single-mass disks}

\begin{figure}
\includegraphics[scale=0.4]{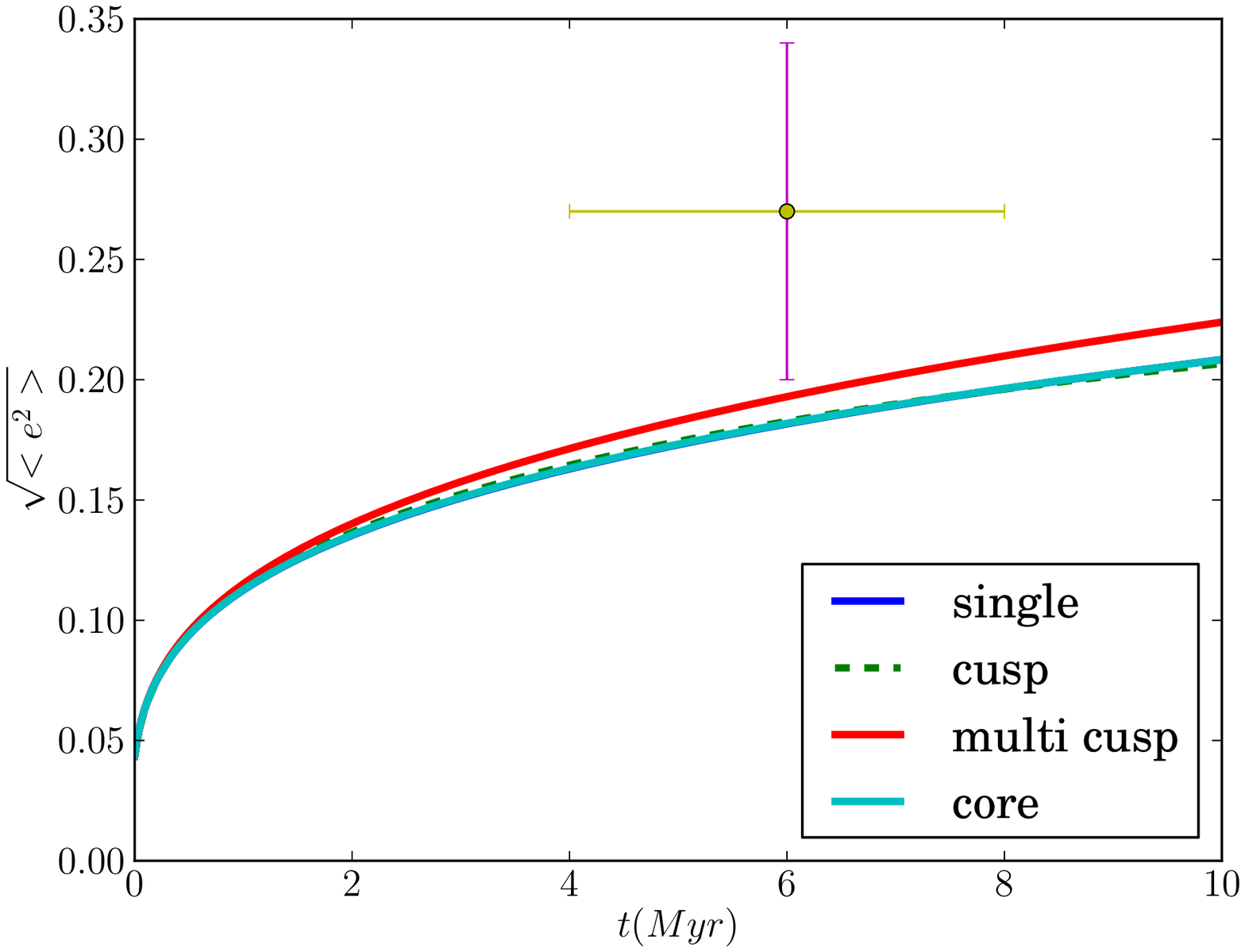}

\includegraphics[scale=0.4]{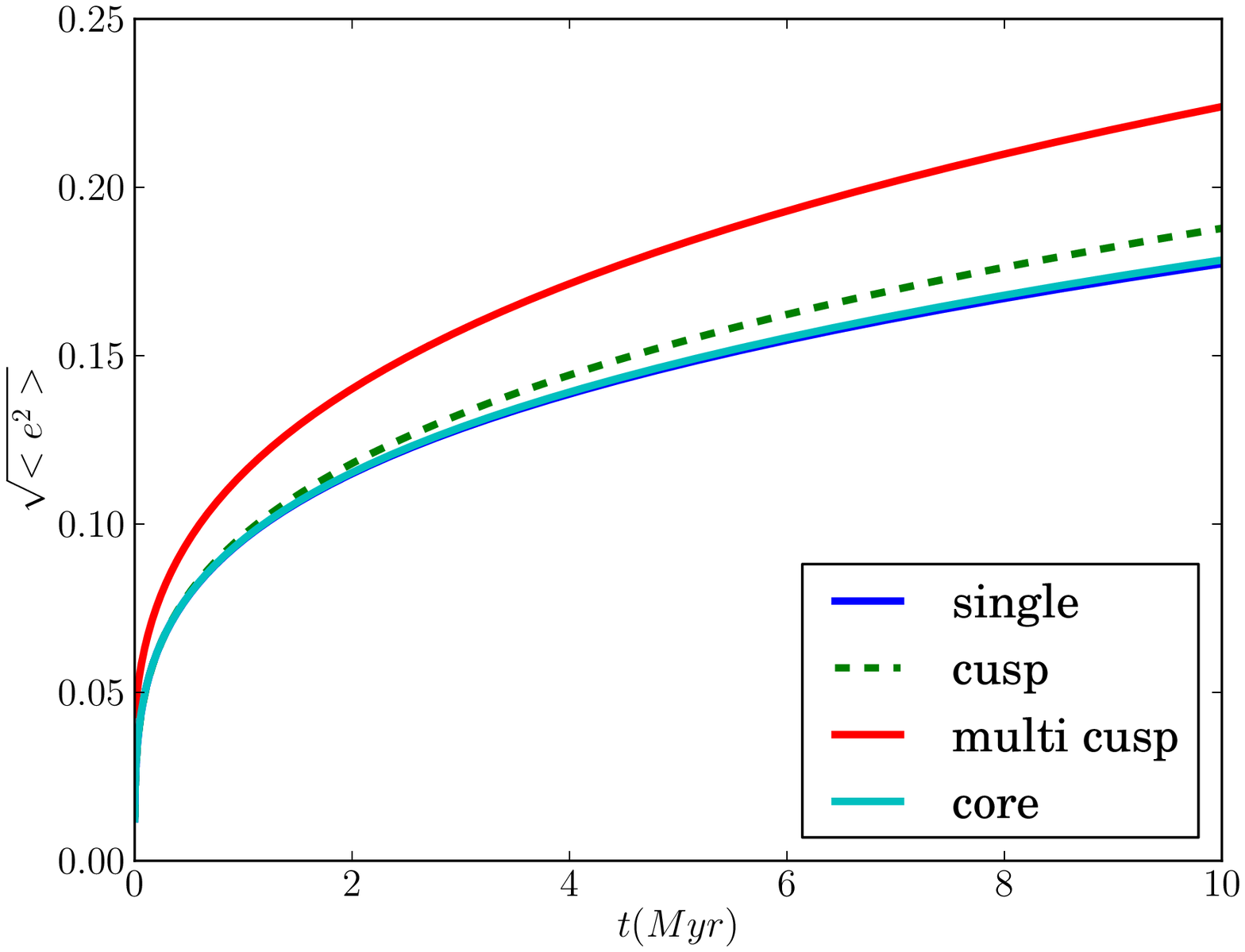}

\caption{\label{fig:single}Evolution of RMS eccentricity of single-mass disk
stars. Upper panel shows models I, II, III and XI. Bottom panel shows
evolution of similar models but with disk stars mass of $10M_{\odot}$.
Also shown are the measured RMS eccentricity of the O-stars in the GC disk\citep{2014ApJ...783..131Y}.}
\end{figure}

Fig. \ref{fig:single} shows the evolution of the RMS eccentricity of the disk stars
for disks with single-mass stars (models I, II, III and XI), for different
choices of mass (10 and 20 $M_{\odot}$ stars). As expected the cusp
heating has a stronger effect on lower mass stars.

\subsection{Dependence on stellar mass and the structure of the nuclear stellar cluster}
\label{sub:Single-Disk:}

\begin{figure}
\includegraphics[scale=0.4]{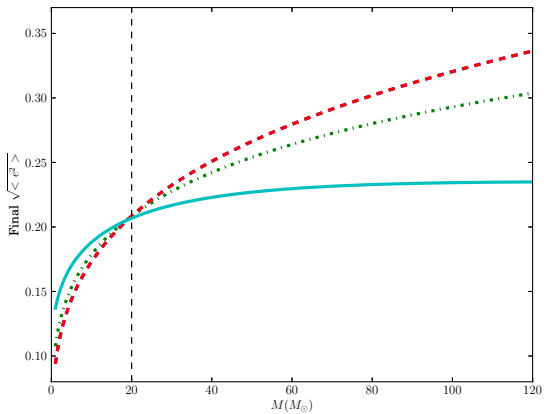} 

\caption{ \label{fig:disk_vs_cusp_core}The RMS eccentricity of disk stars after
$10\,\mbox{Myr}$ of evolution. The solid line corresponds to Model
II, dot-dashed is for core Model III, and the dashed line corresponds
to the case where no NSC exists.}
\end{figure}

Fig. \ref{fig:disk_vs_cusp_core} shows the role of the cusp and its
structure on the evolution of disk stars, and the role of the single-star
masses. The final RMS eccentricity (after $10\mbox{Myr}$) is shown for a range
of single-mass disks (models I, II and III). Each point in
the lines corresponds to a disk with the same total mass, but the mass of the stars  composing the disk,
indicated on the X-axis (same for all stars in the disk) differes. 
Disks composed of lighter stars can be seen to be more sensitive to the heating by NSC
stars, with the cusp model being more effective than the core one. In
contrast, disks composed of heavy stars are cooled down by the NSC
(compared with the model with no existing NSC). 

\subsection{Multi-component NSCs}

\begin{figure}
\includegraphics[scale=0.4]{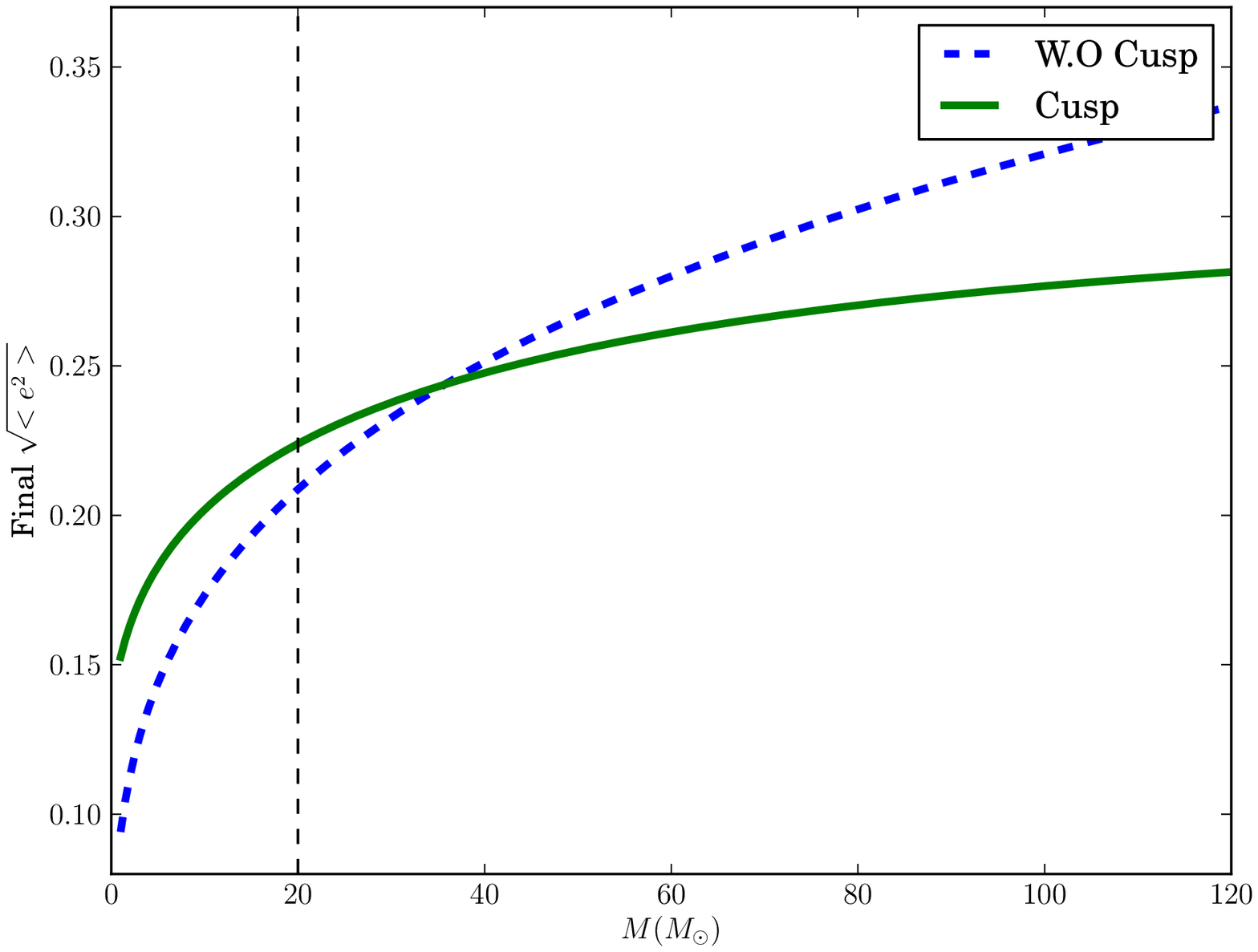} \includegraphics[scale=0.4]{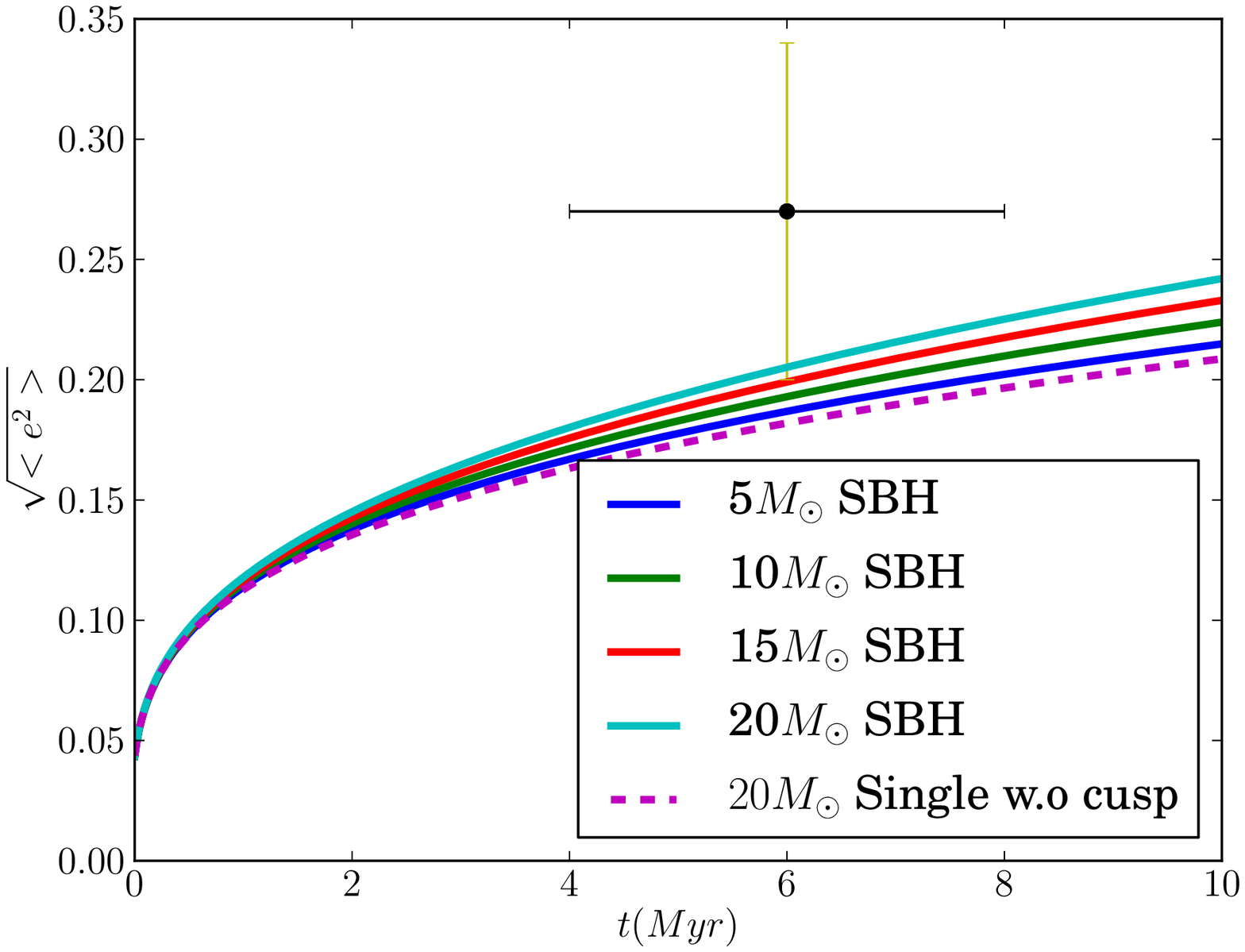}

\caption{\label{fig:Multi_cusp} Evolution of a stellar disk composed of 20
M$_{\odot}$ stars, embedded in a multi-component cuspy NSC (model
XI). The left panels shows the comparison between the disk+multi-component
cusp model and a model without an NSC component. The right panel shows
the evolution of a disk embedded in cuspy NSCs, where different masses
for the NSC SBH population are considered. Also shown are the measured
RMS eccentricity of the GC O-stars \citep{2014ApJ...783..131Y}. }
\end{figure}

In the previous section we studied the evolution of stellar disks
and their interaction with single-mass NSC stellar population. In
the following we consider a more complex and potentially more realistic
model for the NSC. In our multi-component NSC we follow model XI,
where the NSC includes both SBHs and MS solar mass stars. Fig. \ref{fig:Multi_cusp}
shows the disk evolution in such model. Though the number of SBHs
in the cusp is relatively small, they can play an important role in
heating a disk of massive 20 M$_{\odot}$stars, where more massive
SBHs lead to stronger heating of the disk. 

\begin{figure}
\includegraphics[scale=0.4]{multi_cusp_all} \includegraphics[scale=0.4]{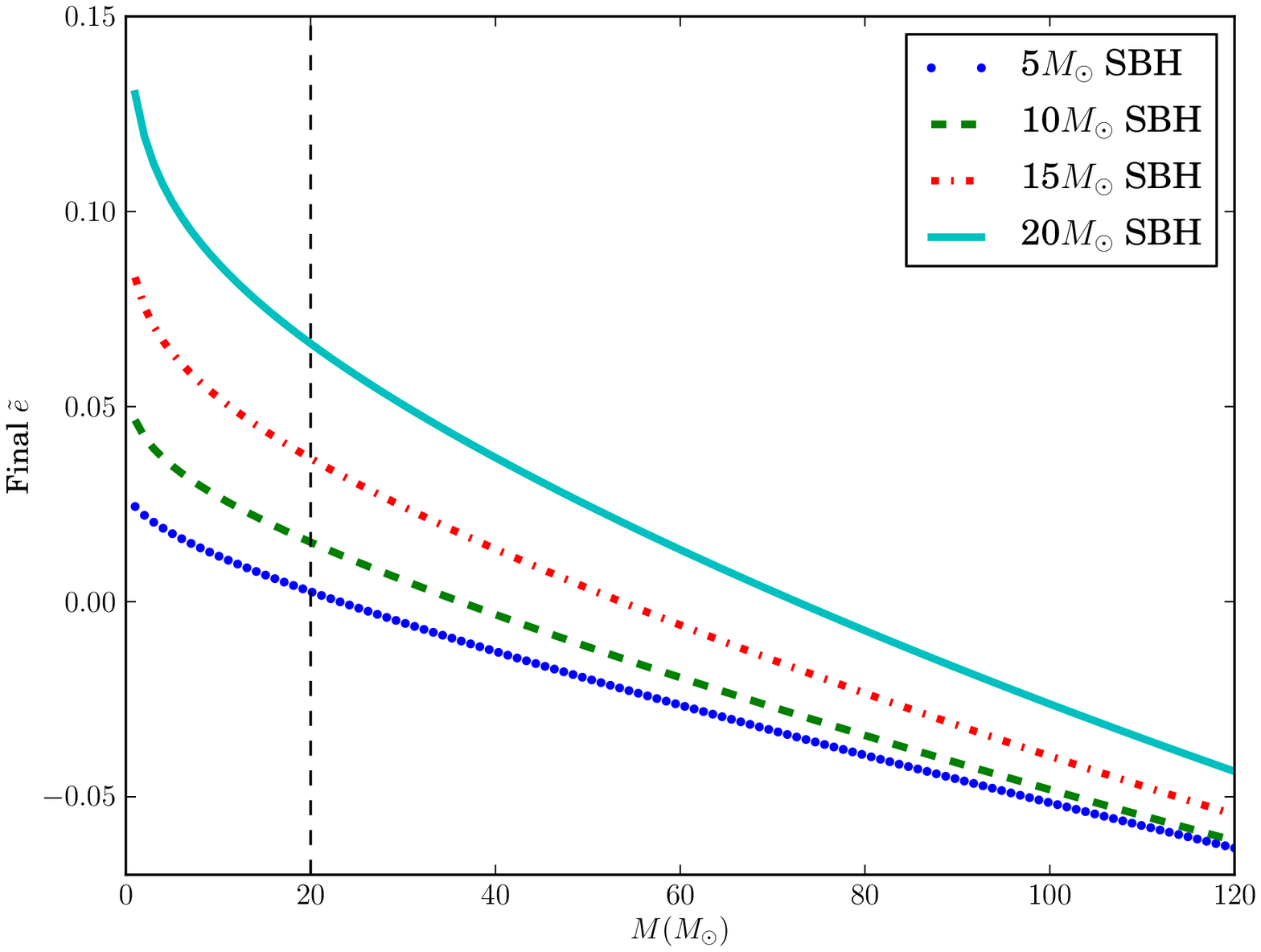}

\caption{Evolution of RMS eccentricity for a disk of O-Stars with multi-component
cusp (model XI), each color is for different SBH mass. The dashed
line is to compare to model I.\label{fig:Multi_cusp-1} On the left
panel is plotted the evolution through the time. The right panel is
the difference between the final eccentricity for each model compared
to model I.}
\end{figure}

Fig. \ref{fig:Multi_cusp-1} shows that the inclusion of SBHs in addition to Solar mass stars in the cusp  (model XI) leads to a more prominent disk heating, as expected.

\subsection{Evolution of multi-mass disks}

\begin{figure*}
\includegraphics[scale=0.4]{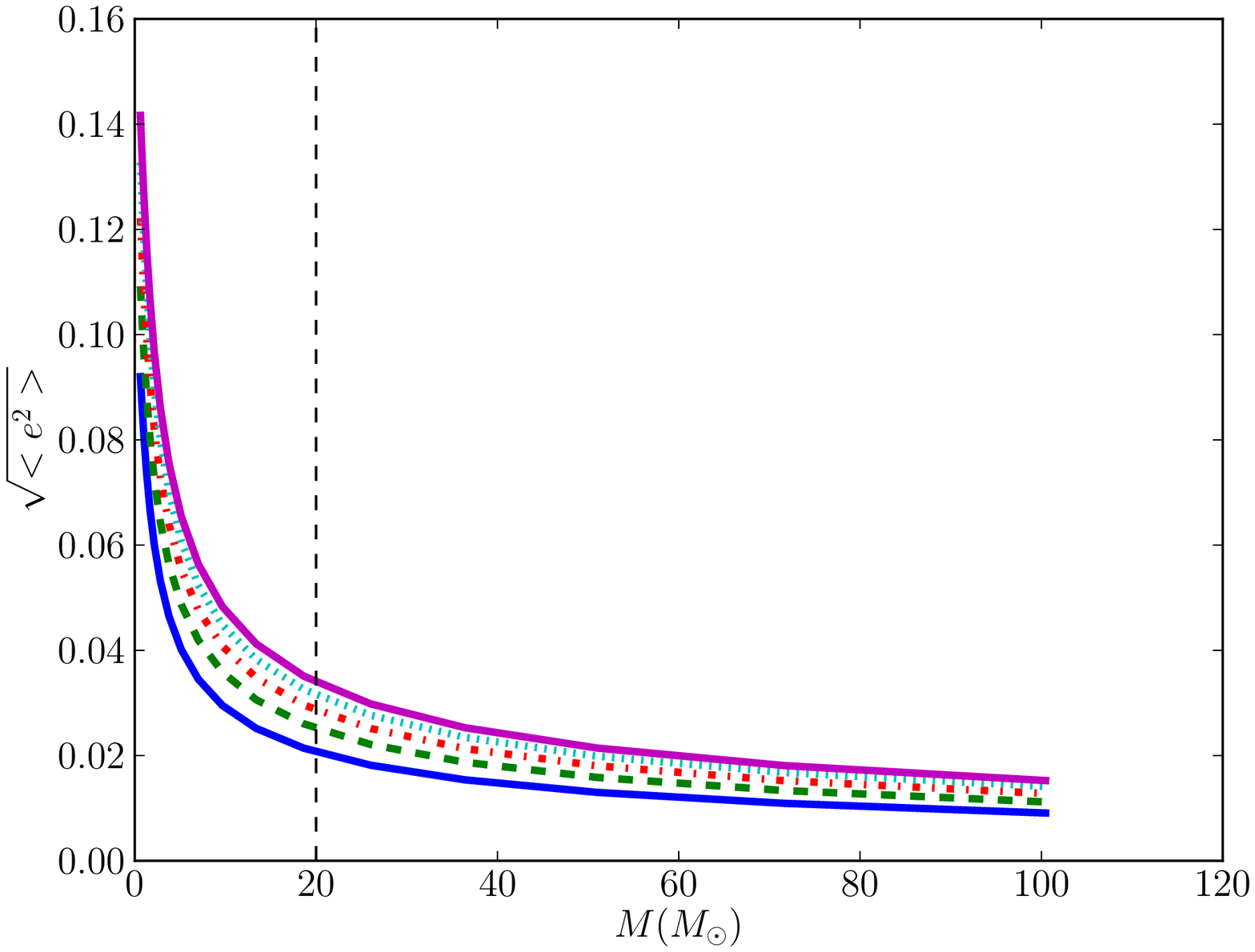}\includegraphics[scale=0.4]{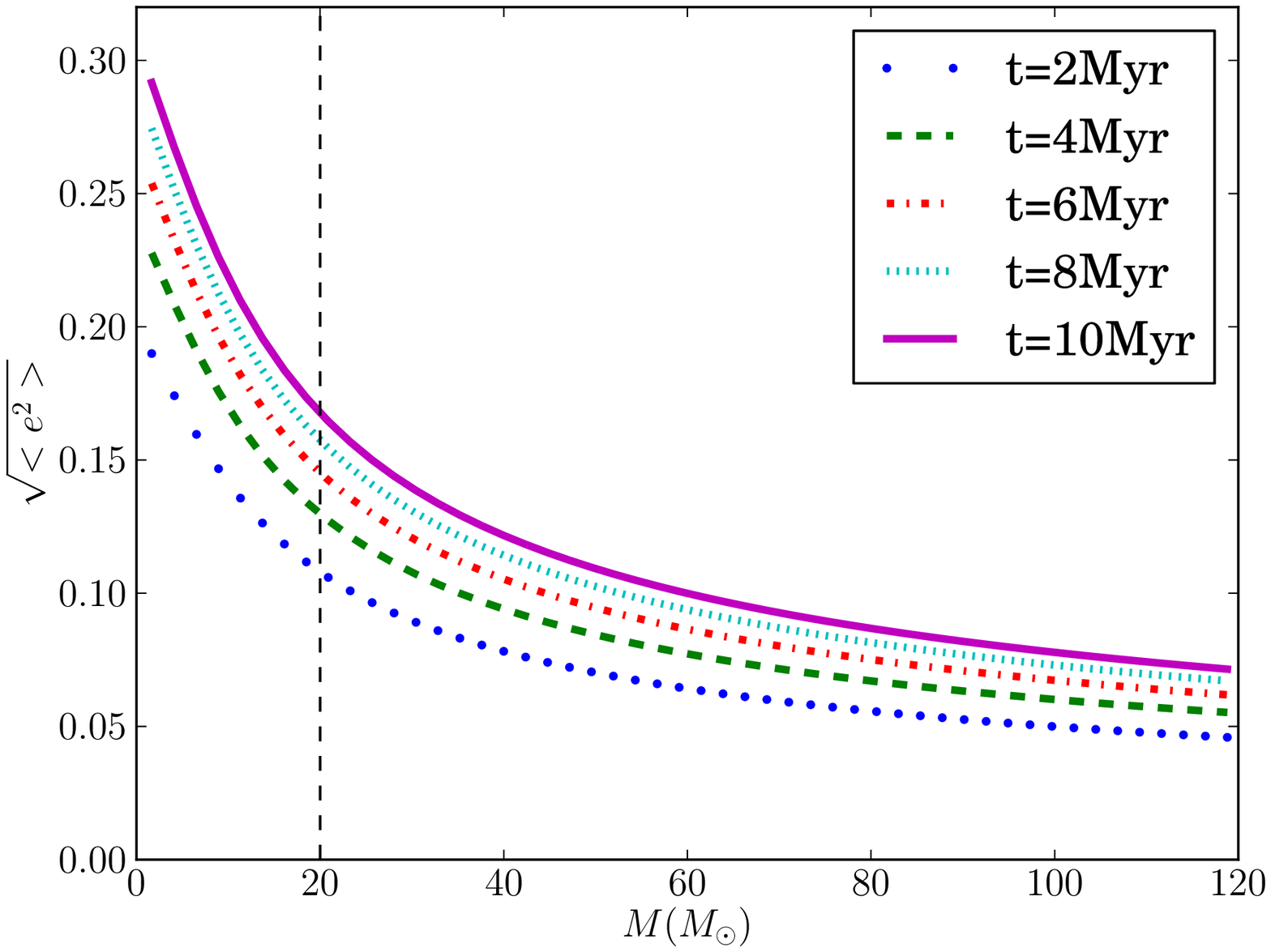}

\caption{\label{fig:IMF} Evolution of the RMS eccentricity of stars with different
masses in the same disk. The upper panel corresponds to a disk with
a Canonical IMF (Salpeter, $\Gamma=2.35$) and the bottom one corresponds
to a disk with a top-heavy IMF ($\Gamma=0.45$). The different lines,
ordered from bottom to top correspond to the evolutionary times: 2,
4, 6, 8 and 10 Myrs, respectively. }
\end{figure*}

\begin{figure*}
\includegraphics[scale=0.4]{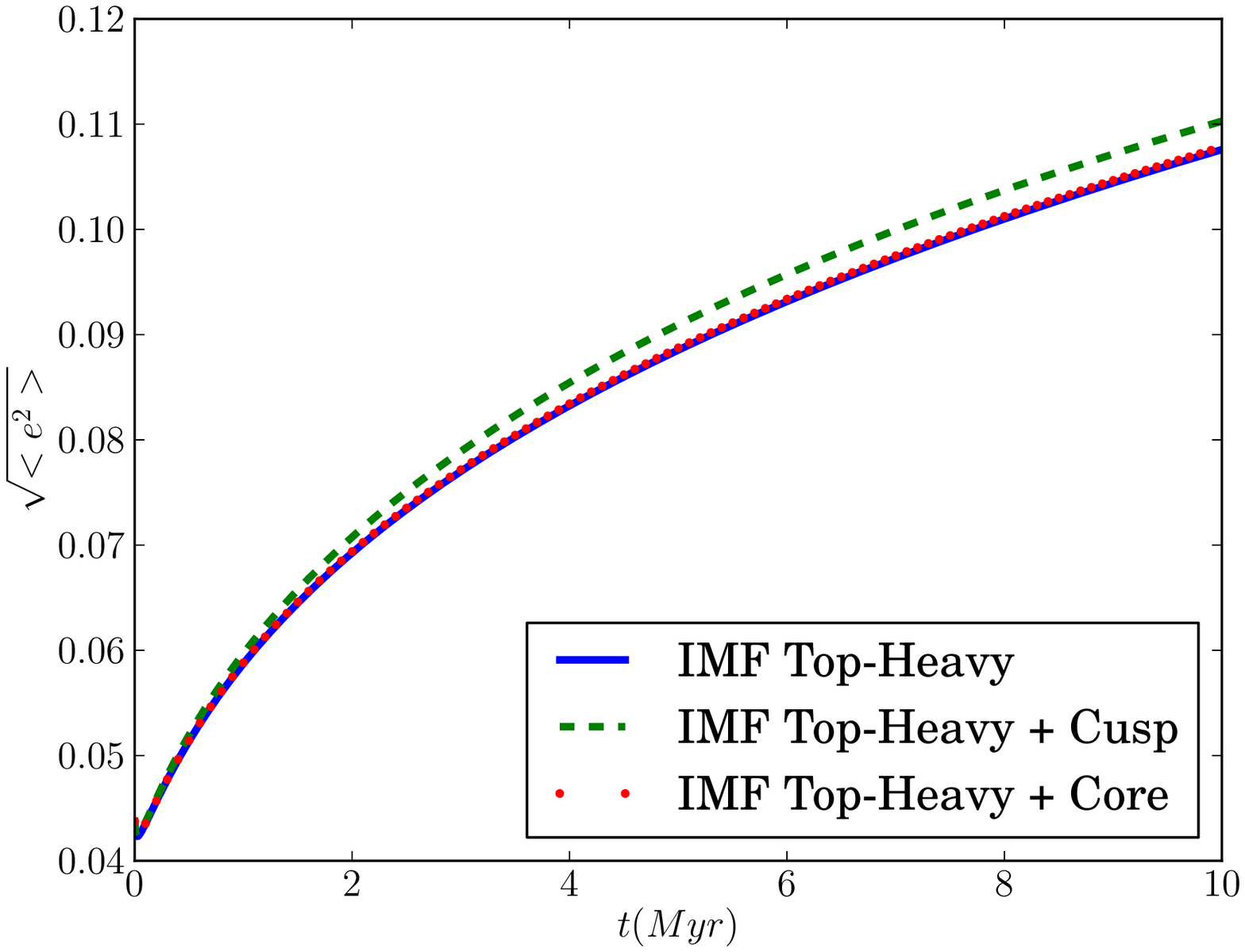}\includegraphics[scale=0.4]{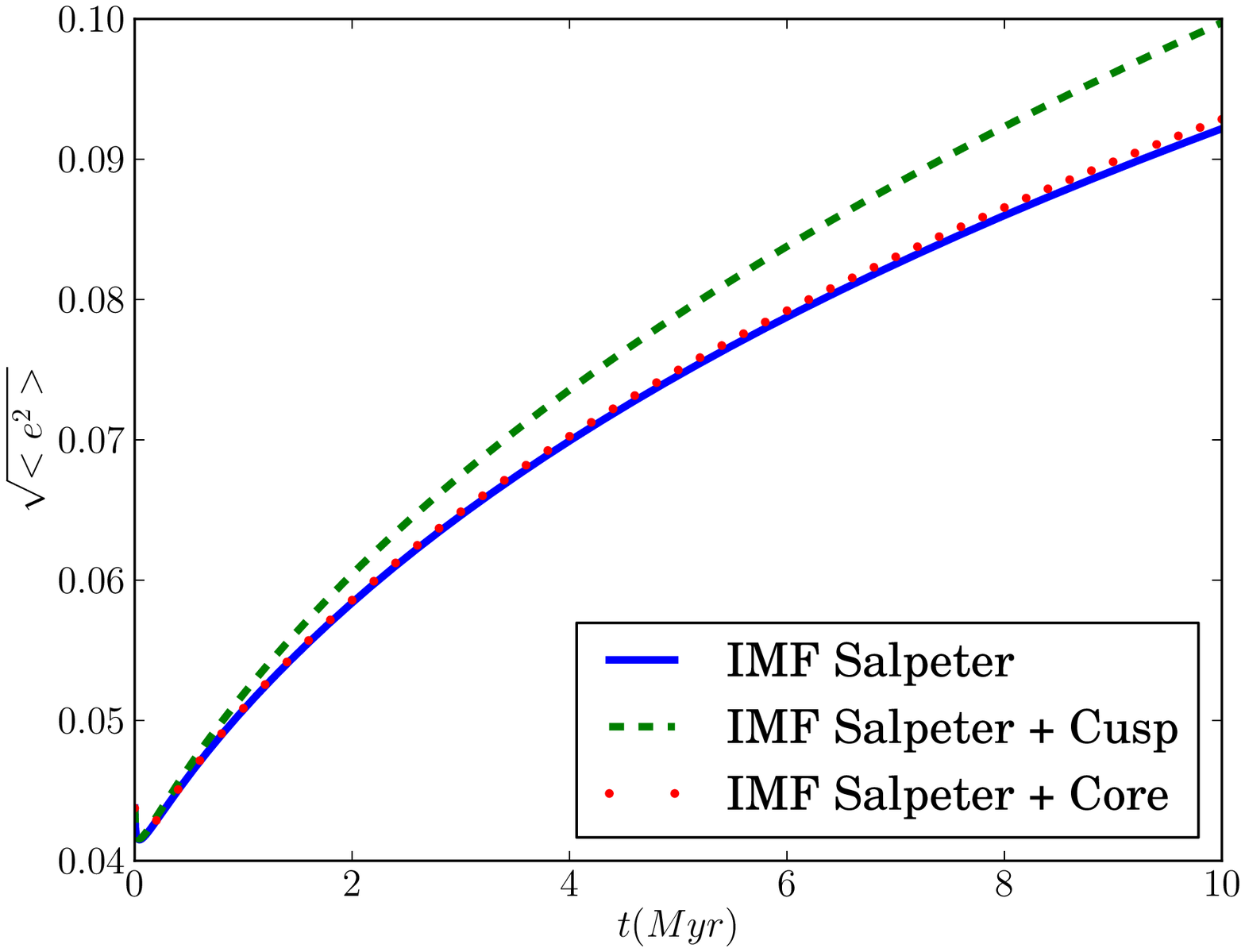}

\caption{\label{fig:IMF_RMS} Evolution of the RMS eccentricity of all stars
(of any mass) in each of the models. The left panel shows top-heavy
IMF ($\Gamma=0.45$) models and the right one shows models with a
Salpeter IMF ($\Gamma=2.35$). Solid lines correspond to models without
any NSC contribution. }
\end{figure*}

In models $IV-IX$ we studied the evolution of multi-mass stellar
populations in a disk, where we considered both disks with Salpeter
IMFs as well as disks with top-heavy IMFs. For each model we divided
the masses into 20 logarithmic mass bins for the Salpeter IMF cases;
for better mass resolution we considered linear mass bins for the
Top-Heavy IMF. For each mass bin we add a coupled equation (\ref{eq:dsigmaj_dt}),
where we took the mass weighted average to represent the mass bin,
with the appropriate number of stars in that mass bin.

Fig. \ref{fig:IMF} shows that the disk evolution leads to a mass
stratified structure of the disk, with lighter stellar populations
being excited to higher eccentricities/inclinations (and larger scale
height for their disk component) compared with populations of more
massive stars, which are segregated to more circular orbits in the
central part of the disk (lower scale height). Such stratification
is also observed for models including the effect of the NSC, where only small differences
exist as a function of the stellar masses; lower mass stars
are more sensitive to the NSC heating (not shown), as expected. 

The top-heavy IMF models include a significantly larger number of
more massive stars, leading to a much faster heating of the stellar
disk, as also observed in the simple models considered by \citet{2007ApJ...654..907A}.
As can be seen in Fig. \ref{fig:IMF_RMS}, the NSC can further contribute
to the disk heating, leading to $\sim10\%$ higher RMS eccentricity, but its contribution
becomes negligible in the top-heavy IMF cases, where the heating is
dominated by the massive stars in the disk. 

In order to compare these results with the measured RMS eccentricity
of the O-stars in the GC we now consider only the RMS eccentricity of 
3-40 M$_{\odot}$ stars, more comparable to the O/B stars observed in the disk today. 
For an age of $5-7$ Myrs for the GC stellar
disk we expect an RMS eccentricity of the order of 0.14, significantly
lower than the measured $\left\langle e\right\rangle =0.27\pm0.07$
\citep{2014ApJ...783..131Y}. 

\begin{figure}
\includegraphics[scale=0.4]{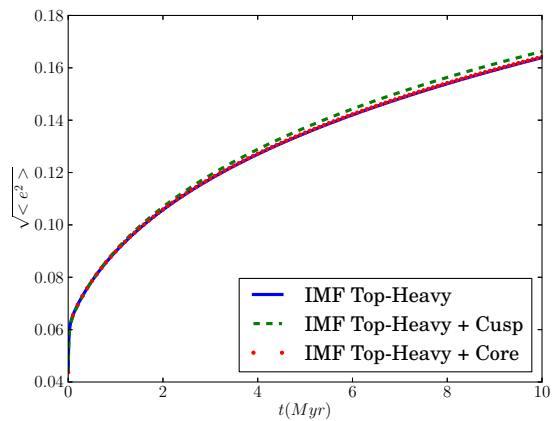}

\caption{\label{fig:RMS-OB-stars} Evolution of the RMS eccentricity of stars
in the mass range $3M_{\odot}<M<40M_{\odot}$ corresponding to OB
stars currently observed in the GC disk. }
\end{figure}

\subsection{Stellar evolution and the long term evolution of stellar disks}

\begin{figure*}
\includegraphics[scale=0.4]{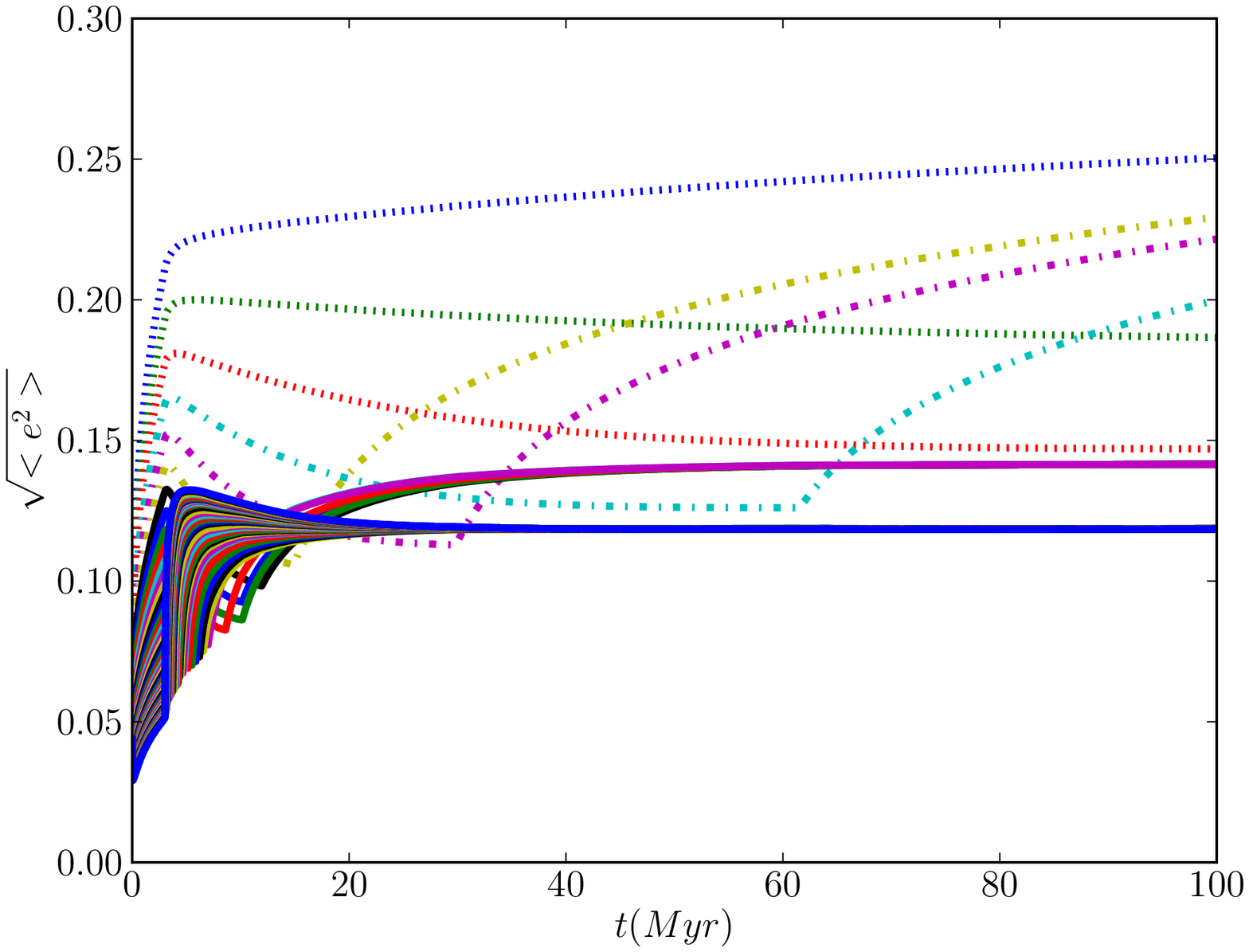}\includegraphics[scale=0.4]{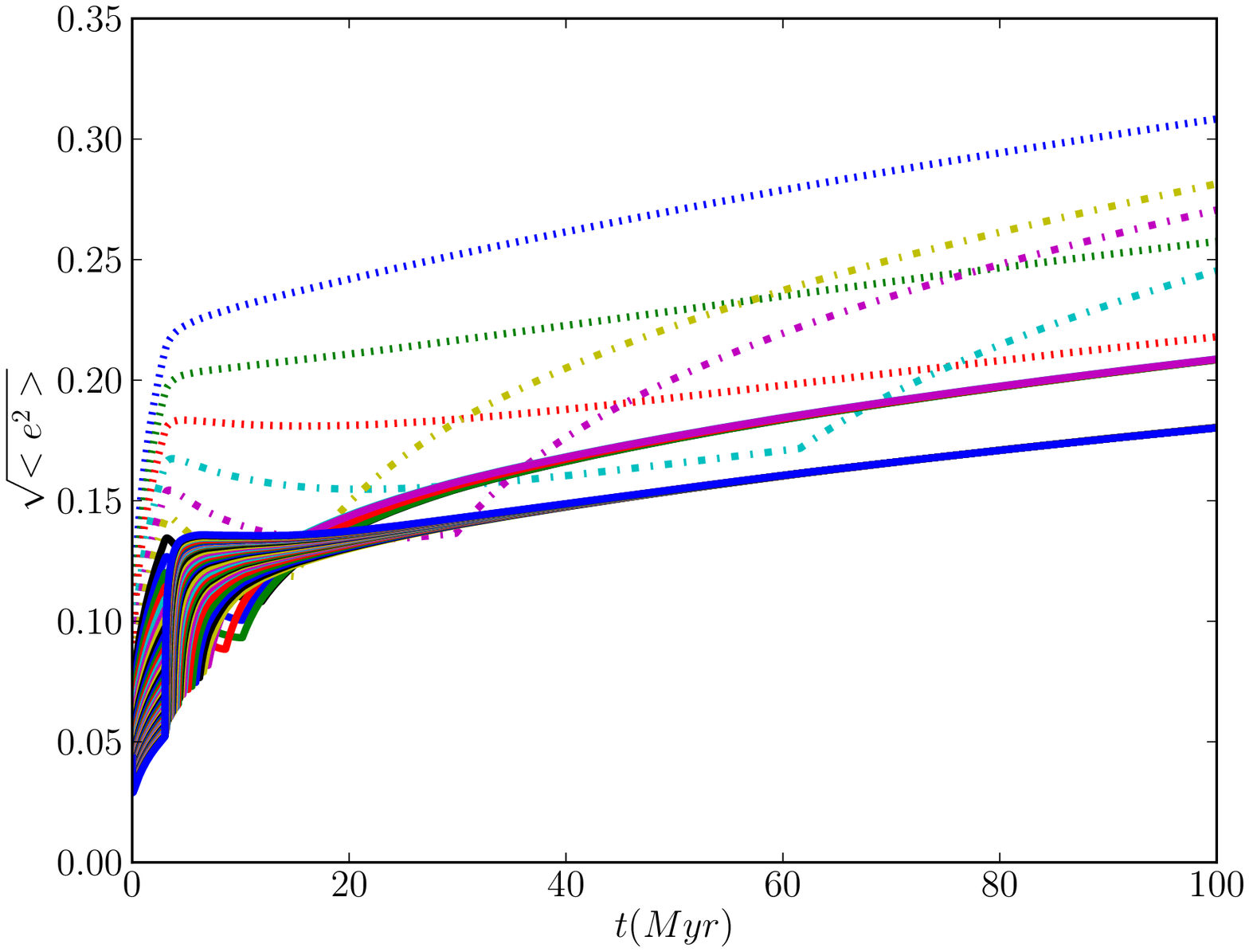}

\includegraphics[scale=0.4]{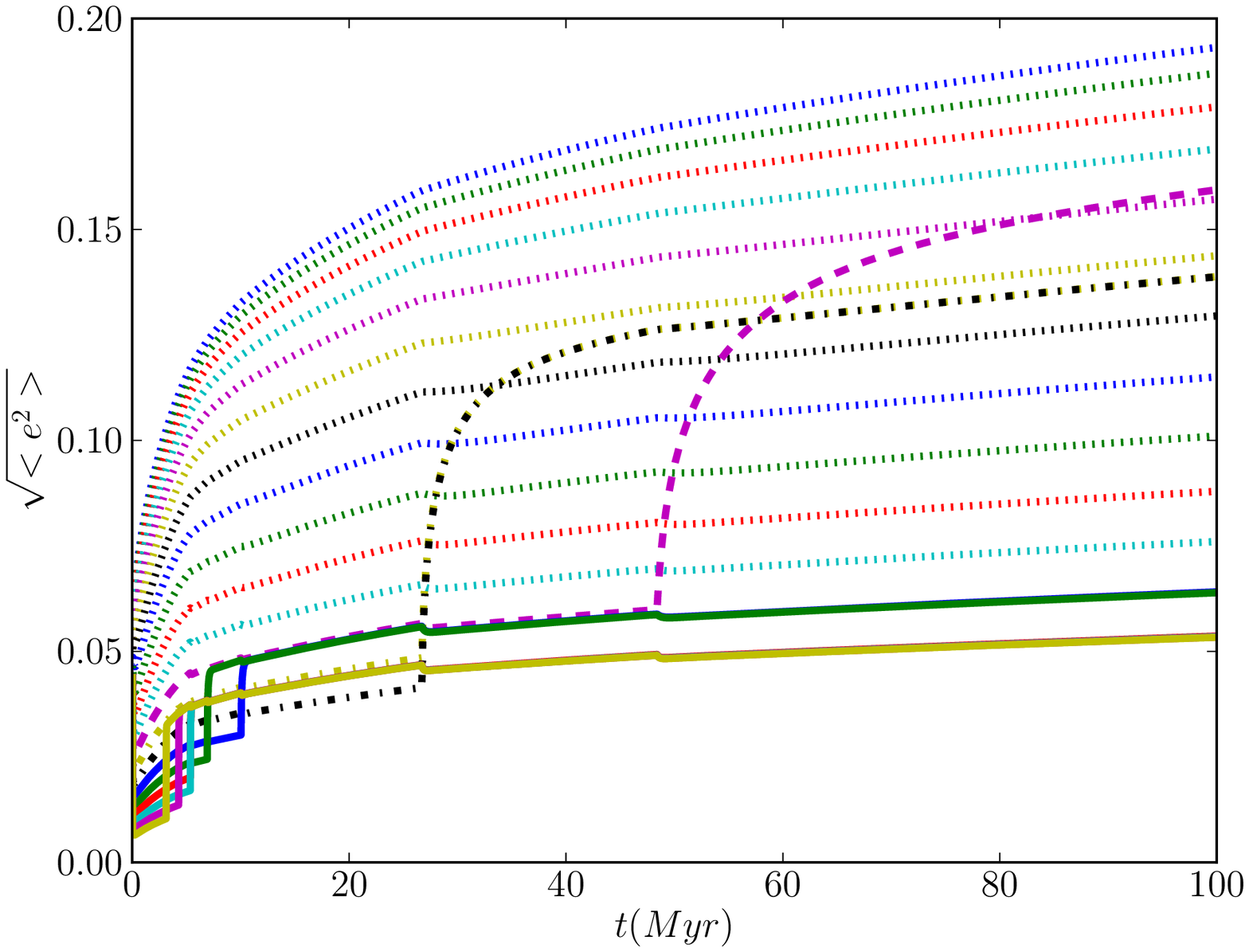}\includegraphics[scale=0.4]{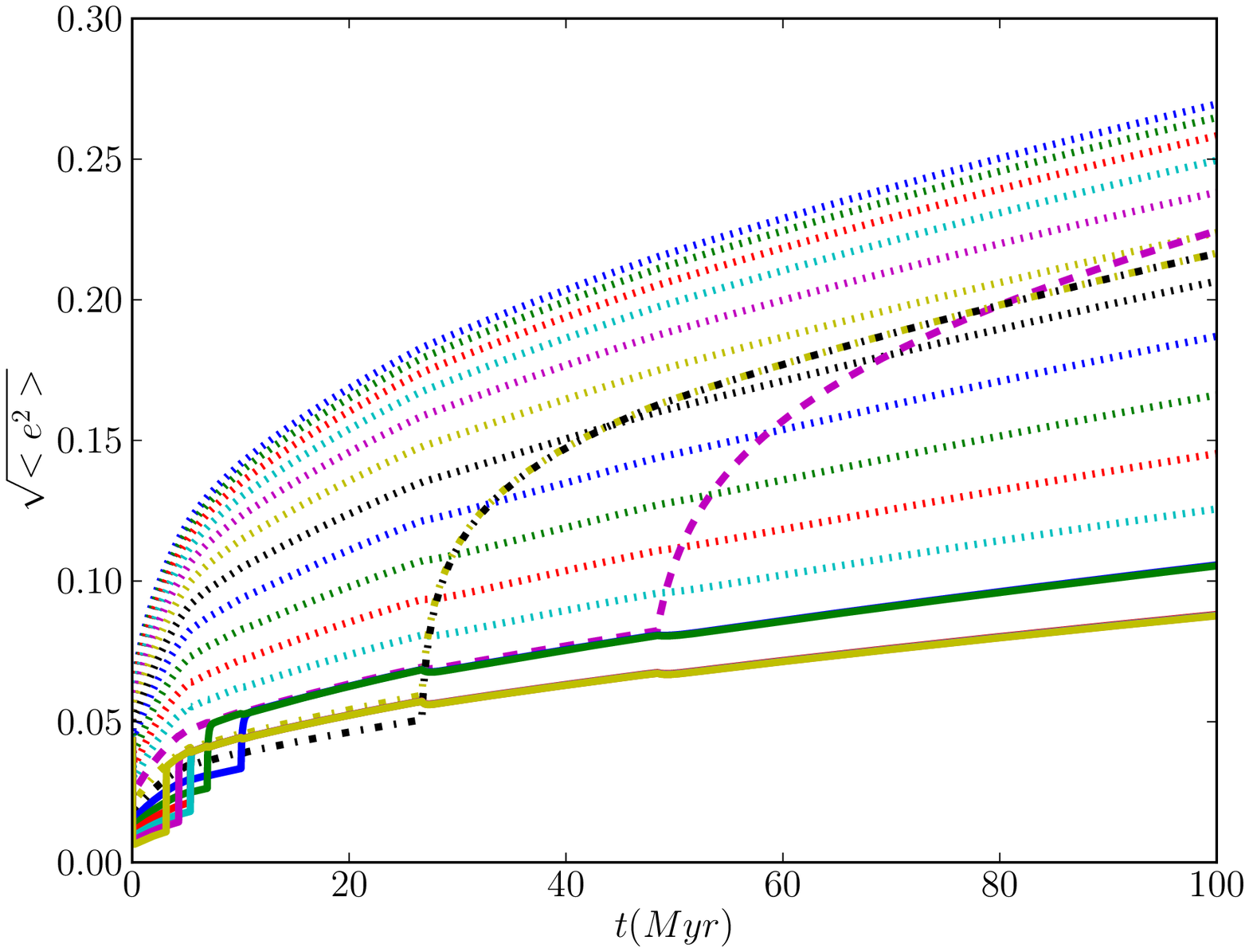}

\caption{\label{fig:long_term_top}The long term evolution of the RMS eccentricity of disk 
stars in over 100 Myrs.  Simplified stellar
evolution is included (see main text). Top panels show the evolution
of a disk with a a Top-Heavy IMF ($\Gamma=0.45$). Bottom panels
show the evolution of a disk with a Salpeter IMF ($\Gamma=0.45$).
The left and right panels correspond to models without an NSC and
with a cuspy NSC, respectively. The lines from bottom to top (initially)
correspond to increasingly less massive stars. The solid, dashed,
dashed-dotted and dotted lines correspond to star that evolve to become SBHs,
NSs, WDs or MSs, respectively.}
\end{figure*}

Fig. \ref{fig:long_term_top} shows
the long term (100 Myr) evolution of the stellar populations in disks
with continuous (Salpeter or top heavy) IMFs, where simplified stellar
evolutionary mass loss is considered (see section \ref{sub:stellar-evolution}). 

As each of the modeled disks evolve, progressively more massive stars
end their life and become lower mass stellar remnants. These (now) lower-mass 
remnants can then be heated much more effectively through
mass-segregation processes. Once the more massive stars evolve the role of the cusp heating
becomes more significant, as be seen from the comparison between the
no-NSC and cuspy NSC models. Moreover, with the absence of high mass
stars intermediate mass populations begin to cool, when no cusp exists;
but these same populations continue to heat up when a cusp population
do exist. 

\section{DISCUSSION}
\label{sec:DISCUSSION:}

In this work we studied the evolution of a stellar disk around a MBH.
We followed previous analytic models developed by \citet{2007ApJ...654..907A},
which considered two-body relaxation and mass segregation processes,
and extended the models to include the effect of binary heating, and
role of the nuclear cluster interactions with the disk, using several
plausible models for the NSC. We also studied realistic mass function
models for the disk stellar populations, and included simplified models
for the stellar evolution of the stars. Finally, we also explored
the long term (100 Myrs) evolution of such disks. In the following
we discuss our results and their implications for the role played
by the different processes and components in sculpting the disk evolution.

\subsection{Disk heating and mass stratification }

Similar to the results of \citet{2007ApJ...654..907A}, we confirm
that for isolated stellar disks, the disk heating is dominated by
dynamical friction /mass-segregation processes in which the more massive
stars heat the low mass stars. Correspondingly, the more massive stars
are cooled by the interaction with the low-mass stars, but two-body
relaxation of the massive stars by themselves still keeps heating
them. Models with top-heavy mass function include a larger fraction
of massive stars, and therefore allow for more rapid heating of the
disk, producing larger RMS eccentricities. 

Our models considered detailed mass-functions for the stars in the
disk, allowing us to estimate the evolution of different stellar populations.
As seen in Fig. \ref{fig:IMF}, the disks develop mass stratification
where higher mass stellar populations are expected to have significantly
lower disk height scale and more circular orbits compared with the
low mass stars in the disk. The existence, or lack of such mass stratification
in the GC stellar disk could therefore be an important tool in assessing
the processes involved in the evolution of the GC disk and its initial
conditions.

\subsection{The role of the cusp}

The effect of the NSC on the evolution of the stellar disk was not
considers before in analytic models. The NSC stellar population has
a high velocity dispersion, i.e. it is an effectively ``hotter''
population than the ``cool'' stellar disk population. The NSC can
therefore potentially heat the stellar disk, even when it is composed
of low-mass stars.

Our results suggest that a dense cuspy NSC can play a non-negligible,
though limited role in heating the stellar disk. When the disk contains
a large population of massive stars, these dominate the disk heating
and evolution, and the NSC plays a relatively minor role. Nevertheless,
even young disks in which massive stars still exist are affected by
the NSC, with top-heavy IMF disks being relatively little affects,
but Salpeter-IMF disks showing non-negligible heating up to $\sim10\%$
higher RMS eccentricities. 

The NSC effects are more prominent in cases where the disk is dominated
by low mass stars. These include unrealistic cases of disks composed
of low single-mass stars, but they also relevant for the long term
evolution of stellar disks with realistic IMFs. In the latter case
stellar evolution leads to the transformation of massive stars into
low mass stellar remnants over time. The effective mass-function of
such evolving stellar disks therefore becomes progressively centered
around lower mass stars and stellar remnants which replace the previously
existing massive stars. Correspondingly, the NSC heating of the cusp
plays a progressively more dominant role in the disk evolution as
can be seen in Fig. \ref{fig:long_term_top}.

\subsection{The role of binary heating}

In this study we extended the analytic study of stellar disk evolution
to include binary heating processes. In principle the binding energy
stored in binaries is significant; however, the rate of energy exchange
between the binaries and the single stars in the disk is slow. Therefore,
binary heating is inefficient in heating the stellar disk, compared
with two-body relaxation and mass-segregation processes. These results
are consistent with N-body simulation results which included binaries
done by \citet{2008MNRAS.388L..64C}. 

Note that we consider the overall averaged evolution of the stellar
disk population; atypical single-binary encounters with short-period
massive binaries can lead to strong kicks ejecting stars at high velocities,
even beyond the escape velocity from the NSC \citep{per+12}. Such
encounters may produce a small number of outlying stars at highly
eccentric/inclined orbits, but are not likely to significantly affect
the overall evolution of the disk and its averaged properties (see
also \citealt{per+08}). 

\subsection{The long term evolution of steallar disks and their signature}
We find that the disk evolution over times as long as 100 Myrs is slow. 
Two-body relaxation is too inefficient for the disk to assimilate into the 
nuclear cluster on such timescales. The disk structure is expected to keep its coherency, 
and be observed as a relatively thin  disk even at 100 Myrs. This suggests that 
the existence of older disks formed before the one currently observed in the GC might still 
be inferred from the stellar kinematics of older, lower mass stars, which may still
show a disk-like struture, unless destroyed/smeared by other non-two-body 
relaxation processes.
 
\subsection{Other physical processes (and caveats)}

Our analytic approach considered the role of two-body relaxation and
binary heating, and included realistic disk and NSC components. However,
secular processes and possible collective effects are beyond the scope
of such a model; we briefly review studies of these processes.

Resonant relaxation processes \citep{1996NewA....1..149R} could potentially
lead to fast evolution of disk stars inclinations through the vector
resonant relaxation (and slower evolution of the eccentricities),
especially in the inner parts of the disk \citep{hop+06}. \citet{koc+11}
suggested that resonant relaxation can produce significant effects,
however they assumed an unrealistic effective mass of NSC stars, which
is 5-10 times larger than expected in the GC (even when including
stellar black holes). Further study of this effect under realistic
conditions could shed more light on this issue.

Fast two-body relaxation as well as secular evolution of disk stars
could also be induced by massive coherent components, massive perturbers
\citep{2007ApJ...656..709P} such as molecular clouds, stellar clusters,
intermediate black hole or a second stellar disk \citep{2009MNRAS.398..429L,yu+07,haa+11,map+13}.
Whether such objects exist/had existed in such configurations as to
influence the evolution of the stellar disk is still actively studied. 

\citet{2009ApJ...697L..44M} considered the collective eccentric disk
instability effect. Such effect could lead to rapid change in the
distribution of eccentricities in the disk, but it requires initially
highly eccentric disk (see also \citealt{gua+12}). Such a process
may give rise to a significant population of highly eccentric stars
in the disk; which are not seen in most recent studies \citep{2014ApJ...783..131Y}.

It is important to note that by their nature, most of the collective
and secular effects mentioned above affect both low mass and high
mass stars in a similar manner, leading to similar eccentricity distribution.
Discerning the kinematic properties of low mass vs. high mass stars
is therefore highly important in order to understand the roles of
such processes compared with stellar two-body relaxation processes
that lead to mass stratification, as shown here in details. In particular,
non-dependence of the stellar kinematics on the stellar masses would
strongly suggest collective effects dominate the stellar disk evolution.

\subsection{Comparison with observations of the Galactic center}

The observed eccentricity of O-stars in the stellar disk in the GC,
$\left\langle e\right\rangle =0.27\pm0.07$ \citep{2014ApJ...783..131Y},
is higher than that obtained in the realistic models considered here.
This suggest either initial conditions with significantly higher eccentricities
for the disk stars, or that other processes besides two-body relaxation
play an important role in the disk evolution. 

Detailed studies of the eccentricity/inclination distribution of disk
stars as a function of mass is not currently available, (but see \citealt{mad+14}
for important global statistical trends). Moreover, lower mass B-stars
in these regions may have a different origin than the disk, and may
have been captured following a binary disruption \citep{2007ApJ...656..709P};
such stars would have high eccentricities \citep{per+10} and may
mask the real distribution of the disk stars eccentricities. Nevertheless,
a focused study on stars most likely related to the disk, and their
eccentricity-mass relation is highly desirable to asses the existence
of possible mass stratification in the stellar disk. 

Finally, two body relaxation processes suggest a factor of two ratio
between the eccentricities and inclination of stars in the disk, while
different relation might be expected from other processes (e.g. secular
Kozai-like evolution induced by an inclined massive structure). A
good handle on the eccentricity vs. inclination from observation could
give additional clues on the dynamical processes involved.

\section{SUMMARY}
\label{sec:SUMMARY}

In this study we analyzed the evolution of a stellar disk around a
massive black hole due to two-body relaxation processes and binary
heating. We explored realistic mass functions for the disk stars,
and included the effects of two-body relaxation by the nuclear cluster
stellar population in which the disk is embedded. We also considered
the effects of stellar evolution and studied the long term evolution
of such disks. The disk evolution is dominated by dynamical friction
from the massive stars in the disk, while binary heating plays only
a negligible role in the disk evolution. The nuclear cluster plays
a minor role in the disk evolution as long as a large population of
massive stars exist, as in the case of a young disk with a top-heavy
mass function; it can play a more significant, though still modest
role in heating a disk with a Salpeter mass function. Due to stellar
evolution the effective mass-function of the disk becomes progressively
centered around lower mass stars and stellar remnants which replace
the previously existing massive stars. Correspondingly, disk heating
by the nuclear cluster plays a progressively more dominant role in
the disk evolution.

We find that significant mass stratification arises from mass-segregation
processes in the disk. This could serve as a signature for two-body
relaxation dominating the disk evolution. In contrast, collective
and secular effects are typically insensitive to the masses of individual
stars in the disk, and would not produce mass stratification. 

The observed RMS eccentricity of the O-stars in the stellar disk in
the Galactic center ($\sim0.27$) is larger than obtained in any of
our models with realistic conditions, suggesting either an initially
hot and/or eccentric disk or that other secular/collective effects
play an important role in the disk evolution. 


\end{document}